\documentclass[11pt]{article}
\usepackage{amsfonts,epsfig,latexsym,amscd}
\usepackage{amssymb}%,letterspace}
\usepackage{amsmath}
\usepackage{mathrsfs}
\usepackage{bbm}
\usepackage[hidelinks]{hyperref}
\usepackage{graphicx}

\usepackage[hang,nooneline]{subfigure}
\usepackage{color}

\newcommand{\R}{{\mathbb{R}}}

\newcommand{\be}{\begin{equation}}
\newcommand{\ee}{\end{equation}}
\newcommand{\bea}{\begin{eqnarray}}
\newcommand{\eea}{\end{eqnarray}}
\newcommand{\bean}{\begin{eqnarray*}}

\newcommand{\eean}{\end{eqnarray*}}
\font\upright=cmu10 scaled\magstep1

\newcommand{\PP}{\hbox{\upright\rlap{I}\kern 1.5pt P}}

\newcommand{\identity}{{\upright\rlap{1}\kern 2.0pt 1}}

\newcommand{\HH}{\mbox{\hbox{\upright\rlap{I}\kern 1.7pt H}}}

\newcommand{\fr}{\frac}
\newcommand{\lm}{\lambda}

\newcommand{\al}{\alpha}

\newcommand{\dg}{\dagger}

\newcommand{\acc}{\\[3mm]}

\usepackage{subfigure}

\def\d{\mathrm{d}}
\def\e{\mathrm{e}}

\def\kihagy#1{}

\def\Tr{\ensuremath{\mathop{\rm tr}}}

\begin{document}
\begin{titlepage}
\strut\hfill
\vspace{0mm}

\begin{center}

{\Large{ \bf Discrete  BPS Skyrmions}}
\\[2mm]
{ \large M. Agaoglou${}^{1,2^{*}}$, E.G. Charalampidis${}^{2^{**}}$,   T.A. Ioannidou${}^{1^{\dg}}$ 
\ and   P.G. Kevrekidis${}^{2^{\ddagger }}$}
\\[12mm]
$^1${\it School of Civil Engineering, Faculty of Engineering,
Aristotle University of Thessaloniki,\newline
 54124 Thessaloniki, Greece }\acc
$^2${\it  Department of Mathematics and Statistics, University of Massachusetts,  Amherst, \newline Massachusetts 01003-4515, USA}
\acc

\end{center}

\vspace{15mm}

\begin{abstract}
A discrete analogue of the  extended Bogomolny-Prasad-Sommerfeld 
(BPS) Skyrme model that admits time-dependent solutions 
is presented. Using the spacing $h$ of adjacent lattice
nodes as a parameter, we identify the spatial profile of the solution 
and the continuation of the relevant branch of solutions over the 
lattice spacing for different values of the potential (free) parameter
$\alpha$. 
%We discuss some of the numerical intricacies of the resulting waveforms including their apparently compact  form which does not allow us in the present setting to carry out a numerical spectral stability analysis. 
In particular, we  explore the dynamics {and stability} 
of the obtained solutions, finding that, while they generally seem to
be prone to instabilities, for suitable values of the lattice spacing 
and for sufficiently large value of $\alpha$, they may be long-lived 
in direct numerical simulations.
\end{abstract}

\vspace{35mm}

$^*${ {\it Email}: {\sf agaoglou@math.umass.edu}} \vspace{-2mm}

$^{**}${ {\it Email}: {\sf charalamp@math.umass.edu}} \vspace{-2mm}

$^\dg${ {\it Email}: {\sf ti3@auth.gr}} \vspace{-2mm}

$^\ddagger${ {\it Email}: {\sf kevrekid@math.umass.edu}}

\end{titlepage}

\section{Introduction}

%Although, the classical Skyrme model \cite{Skyrme} is a good 
%candidate for describing nucleons is unable to describe accurately
%the small binding energy in the nuclei. 

{ The classical Skyrme model~\cite{Skyrme} is a good candidate for 
describing nucleons, although it is unable to describe accurately
the small binding energy in the nuclei}. 
For that reason, {\it generalized} Skyrme models \cite{Ad} that saturate 
the Bogomolny bound have been studied extensively since their mass is roughly
proportional to the baryon number. Recently in \cite{IL}, a submodel of 
the generalized Skyrme model has been considered, which consists, only, 
of the square of the baryon current and a potential term. This model is 
called the BPS Skyrme model since a Bogomolny bound exists and a static 
solution saturates it. 

In the present work, following up on the earlier continuum
work on the BPS Skyrme model of one of the authors~\cite{IL},
as well as the consideration by three of the present authors
of the discrete analogue of the standard Skyrme model~\cite{CIK}, we
embark on an effort to explore the discrete analogue of
the BPS Skyrme model. Upon setting up the relevant formulation,
paying special attention at the domain boundary, we use
numerical bifurcation theory tools to identify the families
of relevant solutions as a function of the lattice spacing
parameter $h$. In this case (differently from what is
the case in the standard Skyrme model), there is an
additional free parameter, namely the exponent $\alpha$ of
the form of the potential energy (cf. for comparison the
continuum case of~\cite{IL}). We utilize similar values
of this exponent as in the continuum case, i.e., $\alpha=4$
and $\alpha=5$, providing a bifurcation analysis in each
case. We observe that in each case, there appears to be
a fold occurring at a finite value of $h$ (in the cases
considered, this value is in the vicinity of $h=0.5$).
Beyond that spacing, no discrete BPS Skyrmions appear
to be accessible. Interestingly, the relevant bifurcation
diagram appears to be significantly different between
the two values of $\alpha$ explored (in one of the two,
the numerical bifurcation curve appears to feature a cusp, while in the
other one, it involves a regular fold).

In both of the above cases, however, a numerical complication
that arises involves the fact that the spatial mode appears to be
nearly compactly supported. A by-product of this, as well as of
the highly nonlinear nature of the model (involving product terms between
the sine of the field and its time-derivative) is the fact
that our attempts to perform a linear stability analysis of such
solutions {in a definitive way} were not successful 
(due to the linearization involving vanishing denominators). 
As a result, we {principally} examined the potential 
stability of the solutions at a fully numerical level involving
direct numerical simulations (DNSs). However, 
we also  performed a suitably tailored stability analysis
(see details below in section 4) for particular
case examples by introducing a regularization avoiding the singular
terms mentioned above. In this 
context, we found that while for the value of $\alpha=4$, solutions
at all lattice spacings considered were found to be unstable, in
the case of $\alpha=5$, long-lived waveforms could be identified,
especially on one side of the relevant fold. These are natural 
candidates for discrete BPS skyrmions, as has been confirmed also
by our stability analysis.

Our presentation is structured as follows. In section 2,
we present the continuum BPS Skyrme model and its energetic 
formulation. In section 3, we turn to the corresponding discrete
model and formulate it dynamically. In section 4, we provide a 
compendium of our numerical results on the latter (discrete) model. 
Finally, in section 5, we summarize our findings and present some
conclusions and challenges for future work.

%%%%%%%%%%%%%%%%%%%%%%%%%
%%%%%%%%%%%%%%%%%%%%%%%%%%

\section{The BPS Skyrme Model}

The action of the BPS Skyrme model is defined by
\begin{equation}
  \label{eq:lag}
  S = \int \d^4 x \left\{
    -\lambda^2 \pi^4 B^\mu B_\mu -\mu^2\, V(U,U^\dag)  \right\},
\end{equation}
 where $U(t,{\bf x})$ is the Skyrme field (that is, an $SU(2)$-valued scalar field);   $\mu$ is a free parameter with units  MeV$^2$;  $V(U,U^\dag)$ is the potential (or mass)  term which breaks the chiral symmetry of the model; $\lm$ is a positive constant with units MeV$^{-1}$;
and $B^\mu$ is the topological current density  defined by
\begin{equation*}
  \label{eq:topcurr}
  B^\mu = \frac{1}{24\pi^2}\,\epsilon^{\mu\nu\rho\sigma}\Tr\left(L_\nu L_\rho L_\sigma\right),
\end{equation*}
where $L_\mu = U^\dagger \partial_\mu U$ is the $su(2)$-valued
current; {to lower and raise indices we use the Minkowski metric tensor $g_{\mu\nu}={\rm diag}(1,-1,-1,-1)$.}

%current and  $g_{\mu\nu}={\rm diag}(1,-1,-1,-1)$ is the Minkowski metric.

By rescaling  $x^\mu \to \left(\lambda n / \sqrt{2}\pi\mu\right)^{1/3}x^\mu$ where $n$ is the baryon number, the action \ (\ref{eq:lag}) becomes
\begin{equation}
  \label{mm}
  S =- \frac{\mu^2}{2}\left( \frac{\lambda n}{\sqrt{2}\pi\mu}\right)^{4/3} \int \d^4 x \left\{
    \frac{1}{144n^2} \left[ \epsilon^{\mu\nu\rho\sigma}\Tr(L_\nu L_\rho L_\sigma)\right]^2 +2 V(U,U^\dag)\right\}.
\end{equation}
In what follows, we consider  the action  (\ref{mm}) rescaled  by   $\left(\lambda n/\sqrt{2}\pi\mu\right)^{4/3}$.

Similarly to the classical case \cite{MI}, we parametrize $U$ by a real scalar field $f$ and a three component unit vector $\hat{\bf n}$ as
\begin{equation*}
  \label{eq:param}
  U = \exp\left( i  f\,  \vec{\bf \sigma}  \cdot \hat{\bf n} \right),
\end{equation*}
where $\vec{\bf \sigma}=(\sigma^1,\sigma^2,\sigma^3)$ are the Pauli matrices.
The unit vector $\bf \hat{n}$
is related to a complex scalar field $\psi$ by the stereographic projection
\begin{equation*}
  \label{eq:param_n}
  \hat{\bf n} = \frac{1}{1+|\psi|^2}\left(\psi + \bar{\psi},-i(\psi-\bar{\psi}), 1-|\psi|^2\right).
\end{equation*}

{For simplicity, spherical symmetry is imposed on $U$ by 
considering a separation of the radial and angular dependence of the 
fields involved in.}
%For simplicity, we consider spherical symmetry. 
%This is done by separation of the radial and  angular dependence of the fields. 
In particular, using the polar coordinates % $(r,\theta,\phi)\in\R^3$ 
{$(r,\theta,\phi)$} we assume that $f=f(r,t)$ and  
$\psi=\psi(\vartheta,\varphi)\equiv \tan\left(\frac{\vartheta}{2}\right) \e^{in\varphi}$.
Then, upon integrating over the azimuthal variables, the action 
(\ref{mm}) assumes the form:
\begin{equation}
  \label{eq:lag_rad}
  S= 2\pi\mu^2\int\d t\int \d r\, r^2 \left[ \frac{\sin^4 f}{r^4}\left( \dot{f}^2 - f'^2 \right) -2 V\right],
\end{equation}
and to get rid of all arbitrary variables we consider the action divided by $\fr{\mu^2}{2}$. Recall that, 
$S=\int L\, dt$  where $L=\int {\cal L}\, r^{2}dr$ and the Lagrangian density ${\cal L}$ is the difference of
%the kinetic and potential terms, that is, ${\cal L}= {\cal T_{\tiny \mbox{BPS}}}-{\cal V_{\tiny \mbox{BPS}}}$.
the kinetic and potential energy densities, that is, ${\cal L}= {\cal T_{\tiny \mbox{BPS}}}-{\cal V_{\tiny \mbox{BPS}}}$.
In what follows, the dot will be used for derivatives
with respect to $t$, while primes for derivatives with respect to the
radial space variable.

Let us concentrate on the  static case, i.e., $f=f(r)$. Then, the energy (\ref{eq:lag_rad})  can be expressed as a sum of a square and a topological quantity. That is,
\begin{equation}
  \label{BPS}
  E =  4\pi\int\d r \,r^2 \left[ \left( \frac{\sin^2 f}{r^2}f' \pm  \sqrt{2 V}\right)^2 \mp 2\frac{\sqrt{2V}\sin^2 f}{r^2}f'\right].
\end{equation}
 
%Since the last term is topologically invariant, in each topological sector
%a minimum is obtained satisfying  the Bogomolny-Prasad-Sommerfield (BPS) equation
{Since the last term is topologically invariant, a minimum can be obtained
for each topological sector satisfying the Bogomolny-Prasad-Sommerfield (BPS) equation}
\begin{equation}
  \label{eq:BPS_eq}
  f' = \mp \frac{\sqrt{2V}r^2}{\sin^2 f}
\end{equation}
 i.e., when the square  term of (\ref{BPS}) vanishes. Solutions of equation (\ref{eq:BPS_eq}) can be obtained either analytically or numerically depending on the form of the potential. 

By choosing a particular type of the potential, the BPS Skyrme model admits topological 
compactons (solitons  with compact support)  which  can be obtained analytically and
reproduce features and properties of the liquid drop model of nuclei \cite{Ad}.
A drawback is that the corresponding time-dependent model does not have a well defined 
Cauchy problem due to the non-standard kinetic term and the non-analytic behaviour of 
the compactons at the boundaries.
{However, the compactons can be transformed into solitons by introducing 
initially a power law potential (with its exponent being treated as a free parameter)
and varying its strength afterward. This way, skyrmions
can be constructed~\cite{IL}.} 
% However, by introducing a  power law (free parameter) at the potential term the compactons
% transforms to solitons by varying the parameter 
% and stable time-dependent skyrmions can be constructed \cite{IL}. 
That is, 
\bea
  \label{eq:pots}
  V_\al &=& \left(1 - \fr{\Tr U}{2} \right)^\al\nonumber\acc
  &=&\left(1-\cos f\right)^\al\nonumber\acc
  &=&\left(1-\sqrt{1-\sin^2 f}\right)^\al\
\eea
where $\al\in \R^+$ is a free parameter. 
Then   for $\al<3$, the solutions are compactons, while for $\al\ge 3$
skyrmion structures can be derived.

\section{The Discrete BPS Skyrme Model}

In this section, a discrete version of the extended BPS 
Skyrme model (based on a Bogomolny-type argument \cite{CIK}) is discussed. 
%In this section,  based on a Bogomolny-type argument presented in \cite{CIK}, a discrete version 
%of the extended BPS Skyrme model is discussed. 
%In order to pass to the setting of the radial lattice,
To embed the current setting into a radial lattice,
$r$ becomes a discrete variable with lattice spacing $h$. So,
the real-valued field $f(r,t)$ depends on the continuum variable
$t$ and the discrete variable $r_{m}\doteq mh$ where $m\in Z^+$. 
Then, $f\doteq f(mh,t)$, $f_+\doteq f((m+1)h,t)$ denotes the forward
shift and thus, the forward difference is given by $\Delta f=(f_+-f)/h$. 
Therefore,  one possibility for discretizing the energy functionals (\ref{BPS}) is to set
\bea
f'&\doteq&\fr{2}{h}\sin\left(\fr{f_+-f}{2}\right),\nonumber\\
\sin f&\doteq&\sin\left(\fr{f_++f}{2}\right).
\label{lat}
\eea
However, the origin should be treated with caution since the energy functionals in (\ref{BPS}) are not defined at $m=0$.
 For that we assume that the Bogomolny bound holds at the origin and therefore, the  energy is given by $E(m=0)=\sqrt{2V}\sin^2f \, f'$  discretized as  (\ref{lat}).
 One can easily check that the discrete version of the
 Bogomolny equation is satisfied especially 
for small $h$. 

Therefore,  the discrete version of the energy is 
\be 
E_{\tiny{ \mbox{dis}}}=4\pi\left[-\sqrt{2^3}\cos^3\fr{f_1}{2}\left(1+\sin\fr{f_1}{2}\right)^{\al/2}+\sum_{m=1}^{\infty}\left( {\cal T}_{\tiny{ \mbox{dis}}}+ {\cal V}_{\tiny{ \mbox{dis}}}\right)\right],
\ee
where the kinetic and potential energy densities are given respectively by
\begin{eqnarray*}
{\cal T}_{\tiny{ \mbox{dis}}}&=& \fr{1}{m^2h} \, \dot{f}^2\sin^4\left(\fr{f_++f}{2}\right),\\
{\cal V}_{\tiny{ \mbox{dis}}}&=&\fr{4}{m^2h^3}\sin^2 \left(\fr{f_+-f}{2}\right)\sin^4\left(\fr{f_++f}{2}\right)+2^{\al+1}m^2h^3
\sin^{2\al}\left(\fr{f_++f}{4}\right).
\end{eqnarray*}

The corresponding  Euler-Lagrange equations take the form
\bea
 && \hspace{-12mm}\fr{2}{h}\sin^4\left(\fr{f_++f}{2}\right)\ddot{f}+\fr{1}{h}\sin^2\left(\fr{f_++f}{2}\right)\sin\left(f_++f\right) \left[2\dot{f}\dot{f_+}+\dot{f}^2\right]=\nonumber
 \acc
 && \hspace{-14mm} -3\,2^{\fr{a-1}{2}}\sin^\al\left(\fr{f+\pi}{4}\right)\,\cos\left(\fr{f}{2}\right)\, \sin f
 +2^{\fr{a-3}{2}}\al \sin^{\al-2}\left(\fr{f+\pi}{4}\right)\,\sin\left(\fr{f+\pi}{2}\right)\cos^3\left(\fr{f}{2}\right)\nonumber\acc
&& \hspace{-14mm} -\fr{2}{h^3}\sin^2 \left(\fr{f_++f}{2}\right)\left[2\sin^2 \left(\fr{f_+-f}{2}\right)\sin \left(f_++f\right)- \sin^2\left(\fr{f_++f}{2}\right)\sin \left(f_+-f\right)\right]
\nonumber\acc
 &&\hspace{-14mm}-2^{\al-1} \al h^3 \sin^{2\al-2}\left(\fr{f_++f}{4}\right)\sin\left(\fr{f_++f}{2}\right),
\ \ \ m=1, \label{eq_m=1}\nonumber \acc
&& \hspace{-14mm}\fr{2}{m^2h}\sin^4\left(\fr{f_++f}{2}\right)\ddot{f}+\fr{1}{m^2h}\sin^2\left(\fr{f_++f}{2}\right)\sin\left(f_++f\right) \left[2\dot{f}\dot{f_+}+\dot{f}^2\right]\nonumber\acc
&& \hspace{-12mm}-\fr{1}{(m-1)^2h}\sin^2\left(\fr{f+f_-}{2}\right)
\sin\left(f+f_-\right)\dot{f}_-^2=\nonumber
 \acc
&& \hspace{-14mm} -\fr{2}{m^2h^3}\sin^2 \left(\fr{f_++f}{2}\right)\left[2\sin^2 \left(\fr{f_+-f}{2}\right)\sin \left(f_++f\right)- \sin^2\left(\fr{f_++f}{2}\right)\sin \left(f_+-f\right)\right]
\nonumber\acc
&& \hspace{-14mm}-\fr{2}{(m-1)^2h^3}\sin^2 \left(\fr{f+f_-}{2}\right)\left[2\sin^2 \left(\fr{f-f_-}{2}\right)\sin \left(f+f_-\right)+\sin^2\left(\fr{f+f_-}{2}\right)\sin \left(f-f_-\right)\right]
 \nonumber\acc
 &&\hspace{-14mm}-2^{\al-1} \al h^3\left[ m^2\sin^{2\al-2}\left(\fr{f_++f}{4}\right)\sin\left(\fr{f_++f}{2}\right)+ (m-1)^2\sin^{2\al-2}\left(\fr{f+f_-}{4}\right)\sin\left(\fr{f+f_-}{2}\right)\right],
\nonumber\acc 
&& \hspace{124mm} m>1 \label{eq_m>1}. \eea
These are the dynamical equations of the model that we will tackle in
the next section numerically.

\section{Numerical Results and Discussion}
% ======================================================================

%
\begin{figure}[!pt]
\begin{center}
\vspace{0.5cm}
\mbox{\hspace{-0.2cm}
\subfigure[][]{\hspace{-0.8cm}
\includegraphics[height=.20\textheight, angle =0]{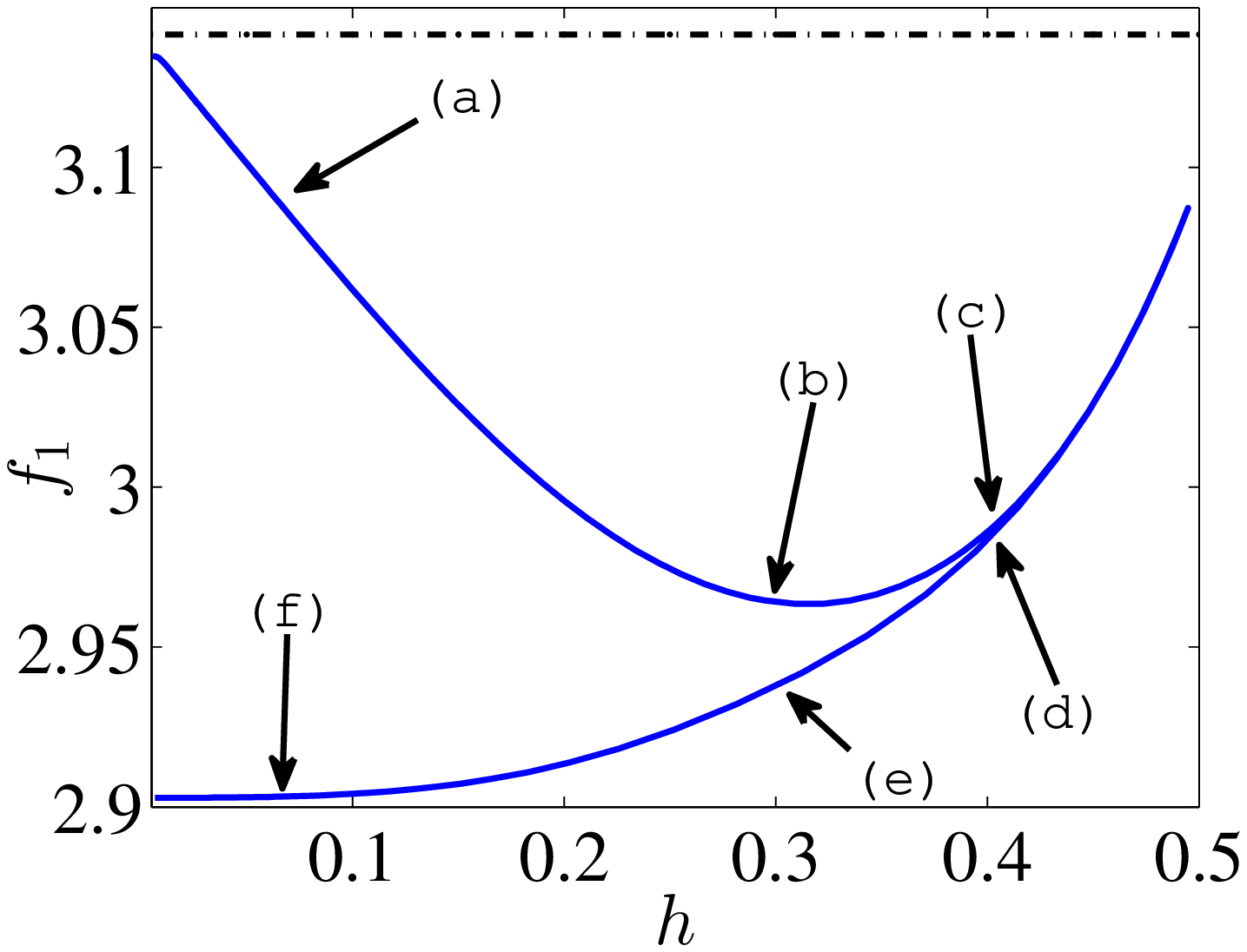}
\label{fig1a}
}
\subfigure[][]{\hspace{-0.5cm}
\includegraphics[height=.20\textheight, angle =0]{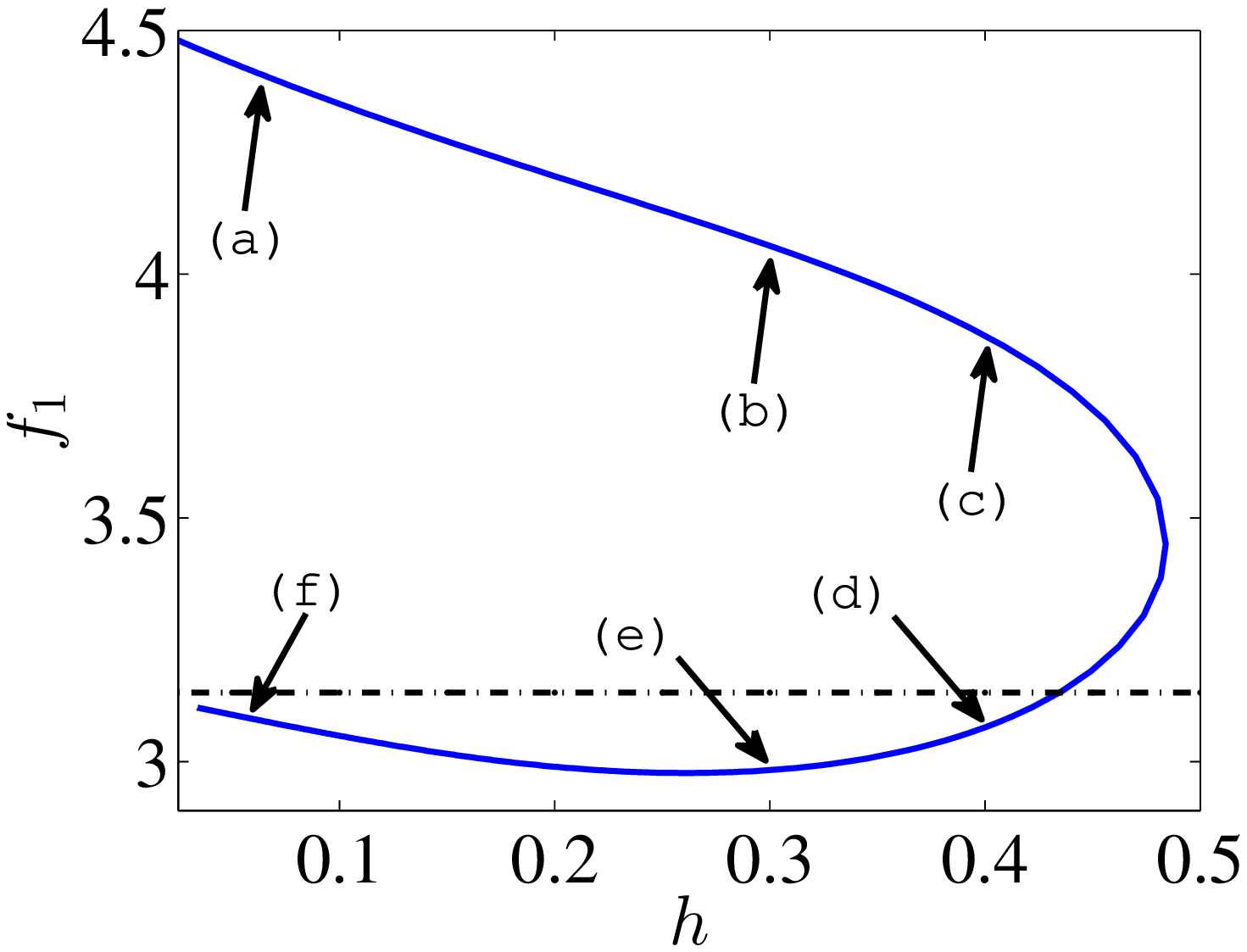}
\label{fig1b}
}
}
\end{center}
\caption{
%Results of continuation over the spacing parameter $h$ for $\alpha=4$ 
%(left panel) and $\alpha=5$ (right panel).
Plot of the static discrete profiles evaluated at the first site of the 
domain as functions of lattice spacing
$h$, i.e., $f_{1}(h)$, for $\alpha=4$ (left panel)
and $\alpha=5$ (right panel). The value of $\pi$ appears in both figures
as a dashed-dotted black line. 
\label{fig1}
}
\end{figure}
%

%In this section, the existence and time evolution of static discrete 
{In this section, the existence, time evolution as well as
(a suitably tailored variant of) stability  of static discrete}
BPS skyrmions are studied when the free parameter $\alpha$ takes the 
values:  $\alpha=4$ and $\alpha=5$. In particular, a Newton-Krylov method
\cite{Kelley_nsoli} has been employed together with a suitable initial 
guess in order to ensure convergence towards a steady-state solution 
(i.e., a stationary BPS skyrmion on the lattice) of equations 
%(\ref{eq_m=1}) and 
(\ref{eq_m>1}).
As a starting point (developing the continuum limit solution),
the BPS equation \eqref{eq:BPS_eq}  with the minus sign
was solved numerically  using a spline
collocation method \cite{COLSYS}.
Then, the obtained profile function $f(r)$ was fed to the Newton solver in order
to identify steady states of the lattice equations
%~\eqref{eq_m=1} 
\eqref{eq_m>1} within $10^{-10}$ (user-prescribed) tolerance.
Finally, a pseudo-arclength continuation over the lattice spacing $h$ was performed by 
utilizing the bifurcation software AUTO \cite{AUTO}. 

%For instance, and as per the cases with $\alpha=4$ and $\alpha=5$, 
%In Figure \ref{fig1}, plots of  the value of the profile
%function at the first site in the domain  $f_{1}$  in terms of  the lattice spacing $h$ is presented  for  $\alpha=4$ and   $\alpha=5$.
%In addition, 
%sample examples of static lattice BPS skyrmions are shown 
%in Figures \ref{fig2} and \ref{fig3} plots of $f_n$ in terms of $n$ are presented for various
%values of $h$. Note that, the captions in the pertaining figures are related to the  the arrows appearing in Figure \ref{fig1}.

\begin{figure}[!pt]
\begin{center}
\vspace{0.5cm}
\mbox{\hspace{0.0cm}
\subfigure[][]{\hspace{-1.0cm}
\includegraphics[height=.18\textheight, angle =0]{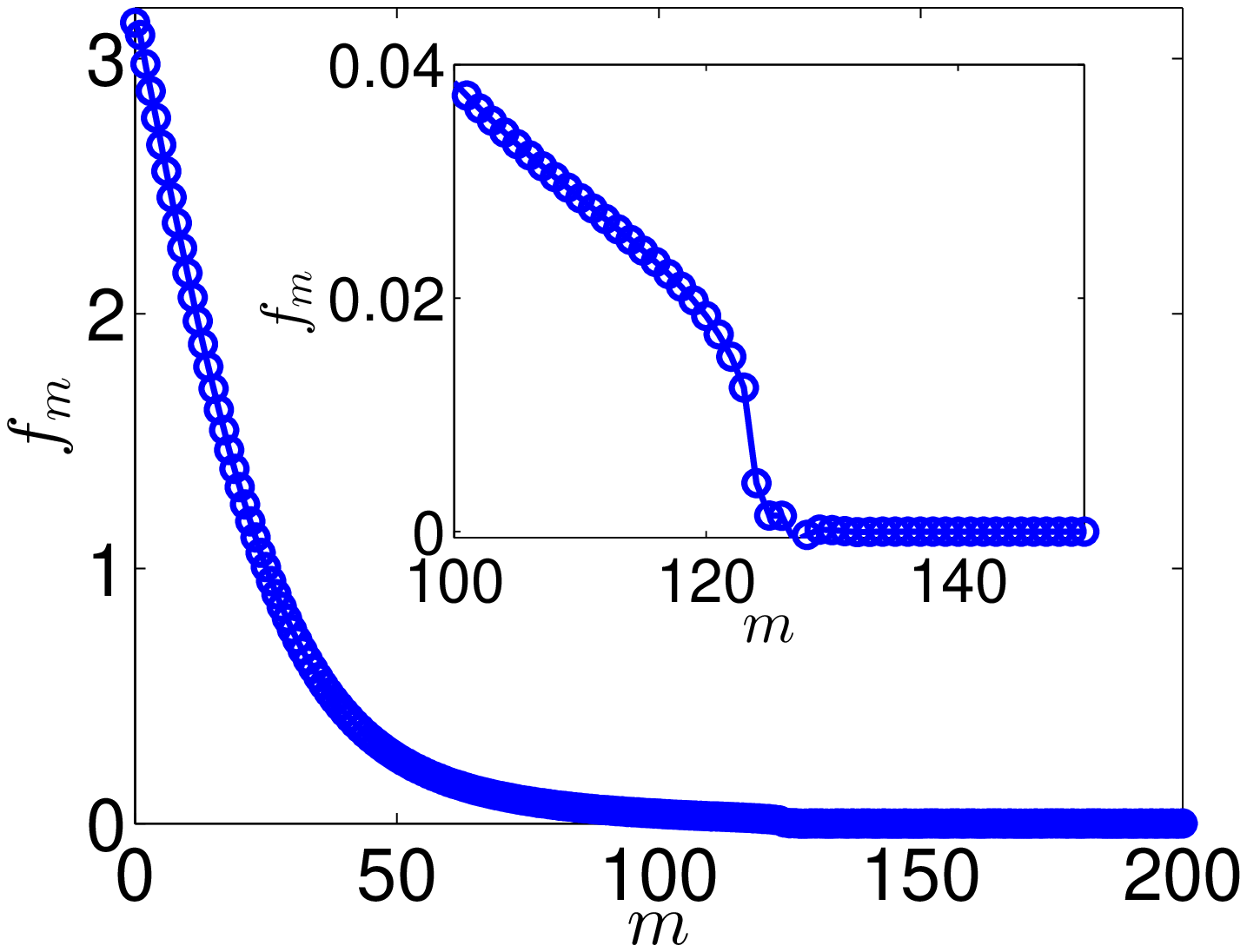}
}
\subfigure[][]{\hspace{-0.5cm}
\includegraphics[height=.18\textheight, angle =0]{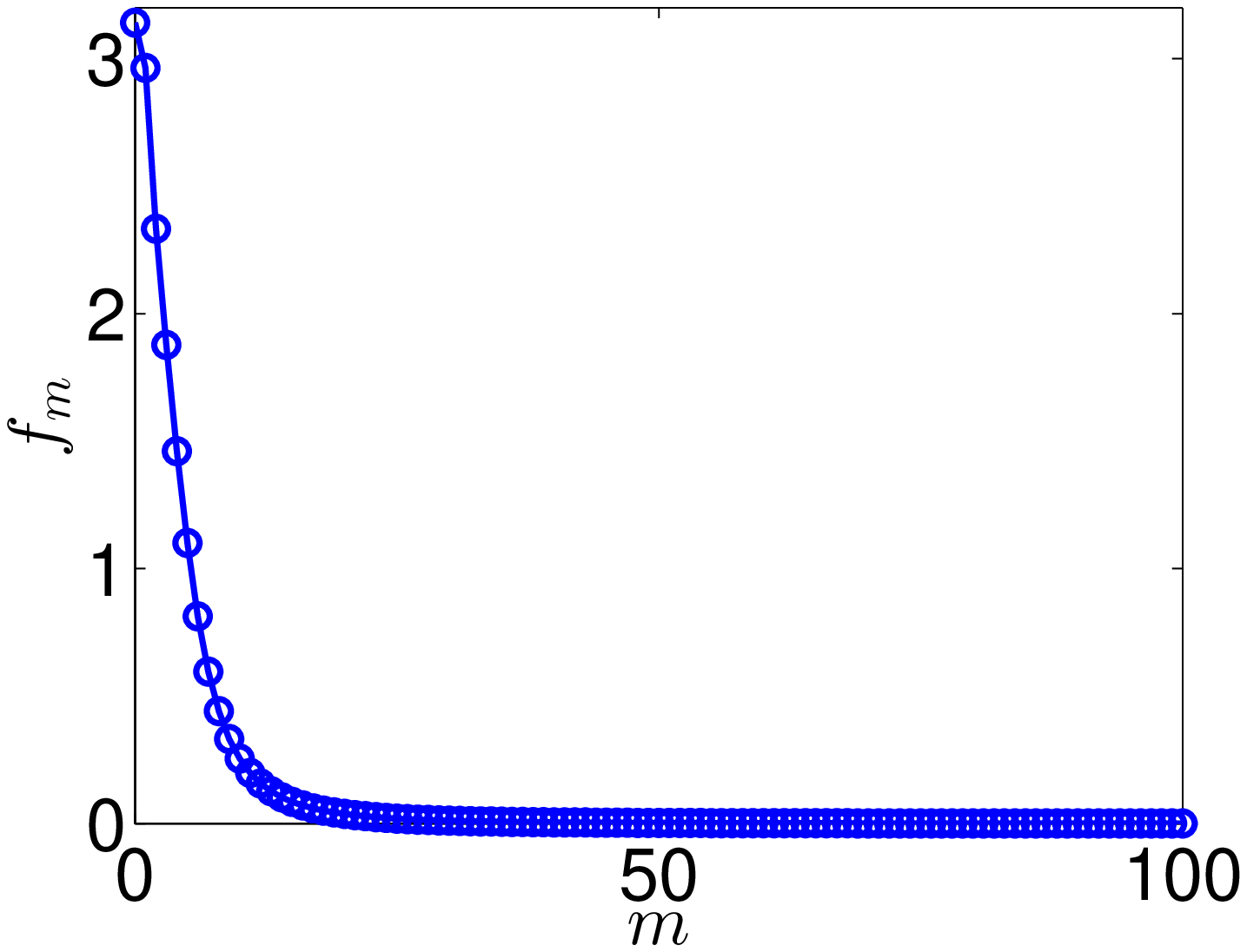}
}
\subfigure[][]{\hspace{-0.5cm}
\includegraphics[height=.18\textheight, angle =0]{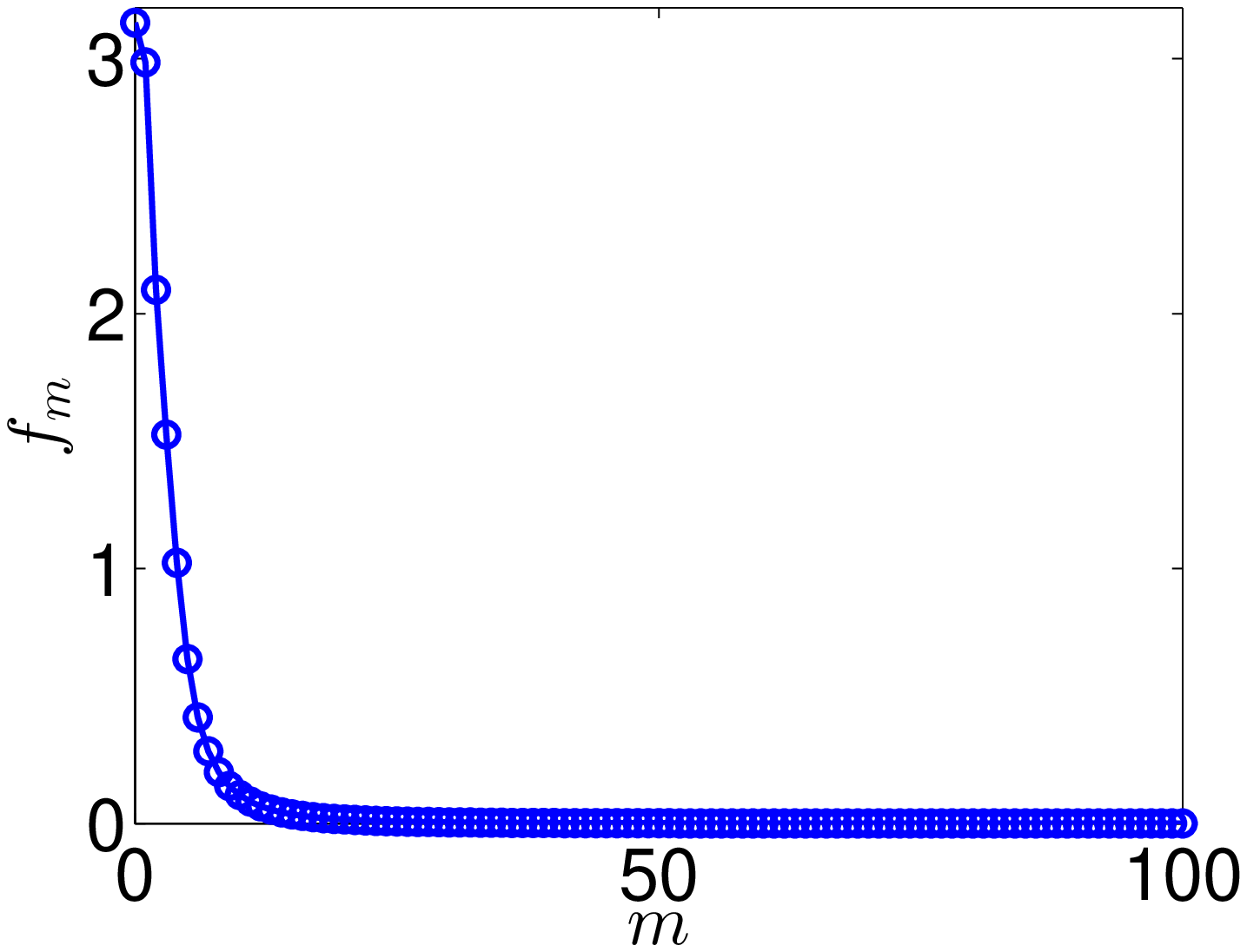}
}
}
\mbox{\hspace{0.0cm}
\subfigure[][]{\hspace{-1.0cm}
\includegraphics[height=.18\textheight, angle =0]{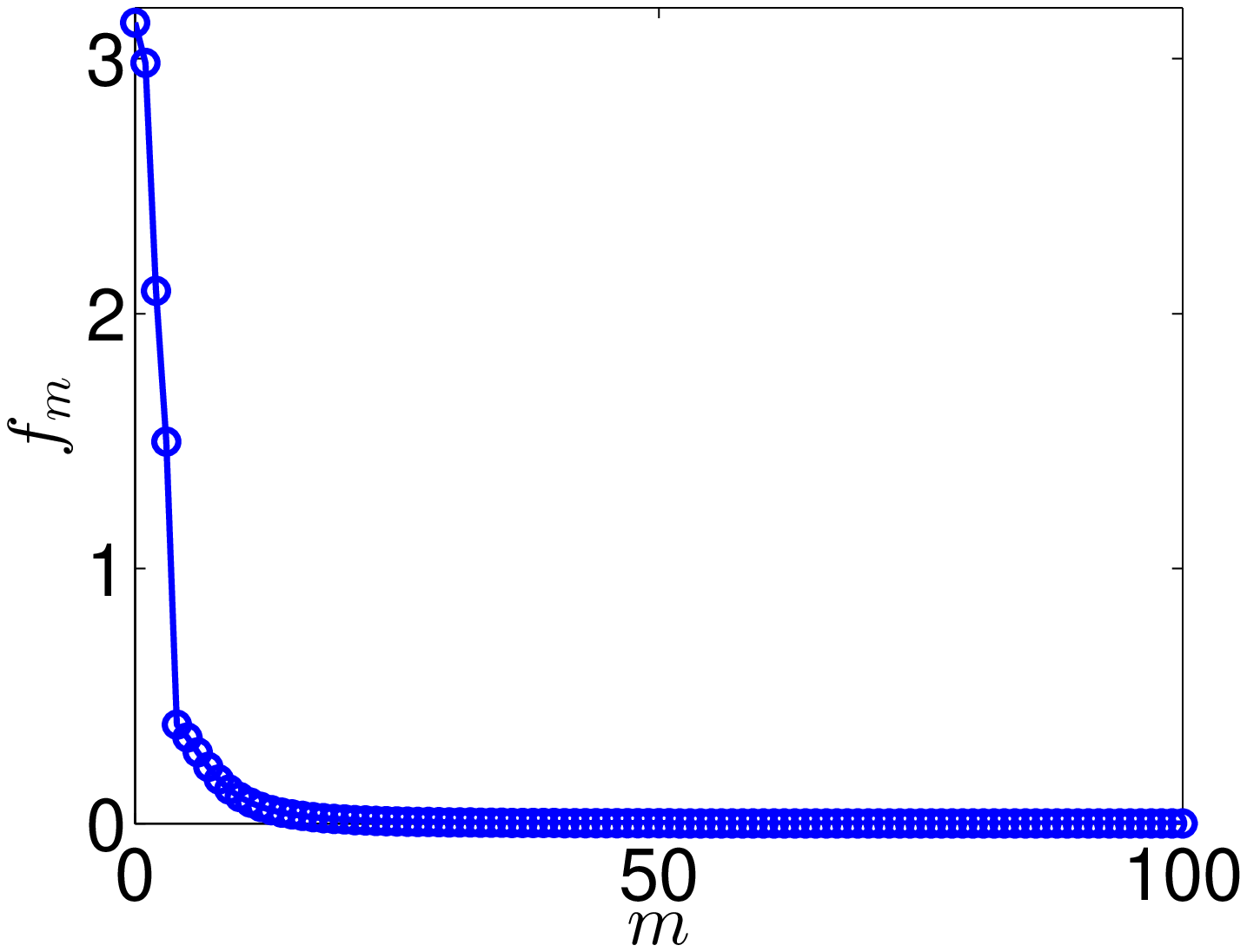}
}
\subfigure[][]{\hspace{-0.5cm}
\includegraphics[height=.18\textheight, angle =0]{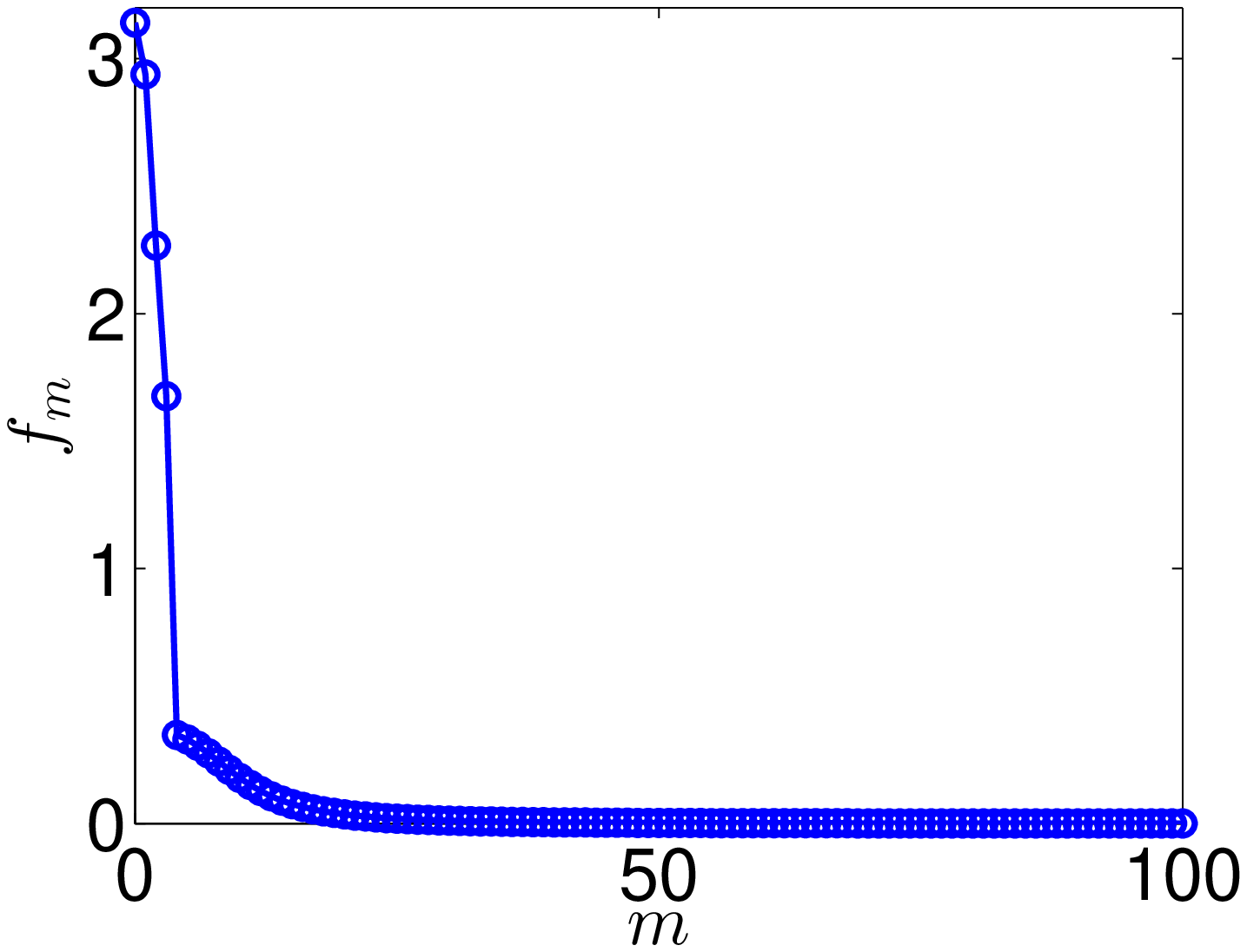}
}
\subfigure[][]{\hspace{-0.5cm}
\includegraphics[height=.18\textheight, angle =0]{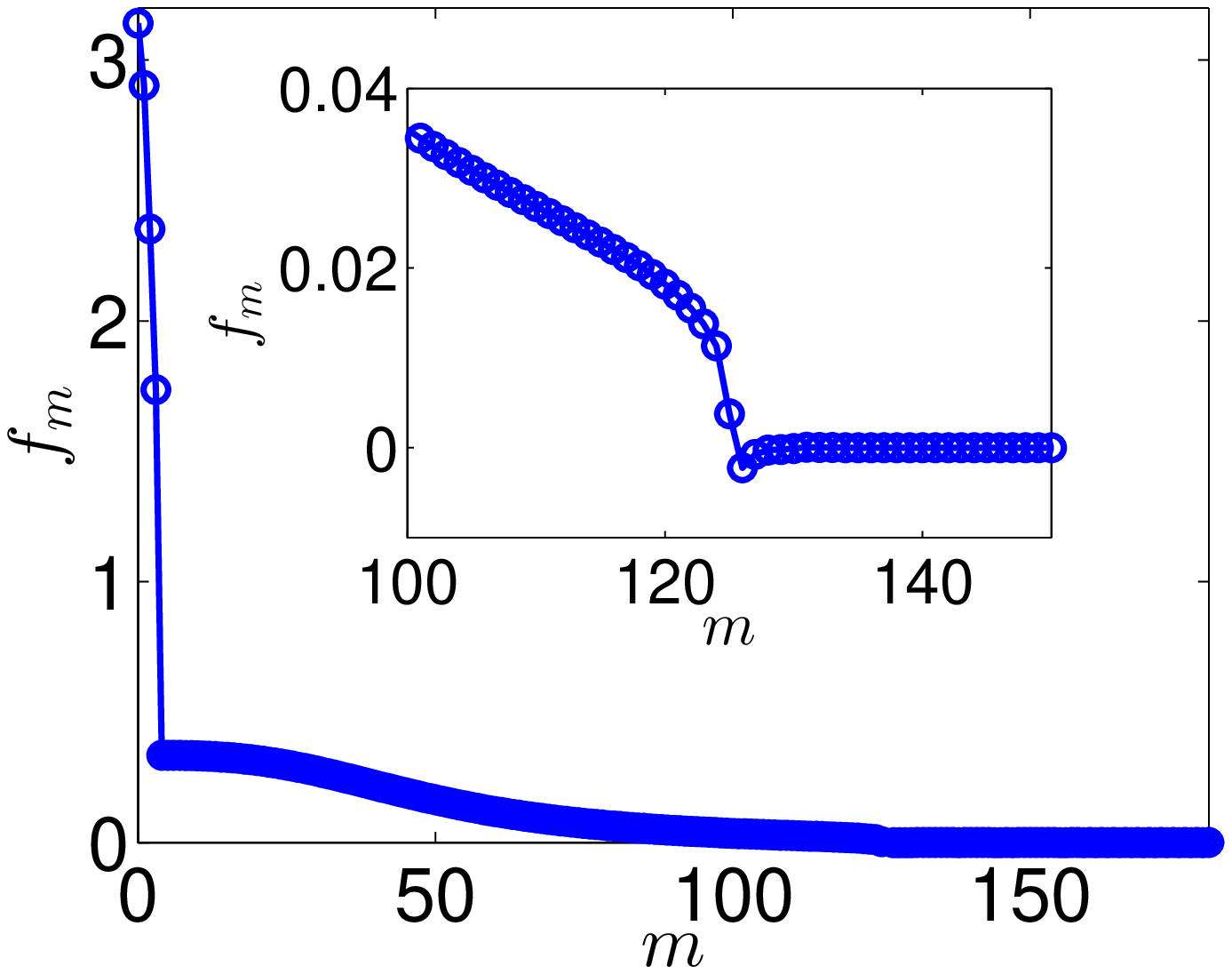}
}
}
\end{center}
\caption{
Static discrete profiles corresponding to the $\alpha=4$ branch 
(see, the left panel of Figure \ref{fig1}) for $h=0.06$: panels (a) and (f);
$h=0.3$: panels (b) and (e); and  $h=0.4$: panels (c) and (d).
\label{fig2}
}
\end{figure}

\begin{figure}[!ht]
\begin{center}
\vspace{0.5cm}
\mbox{\hspace{0.0cm}
\subfigure[][]{\hspace{-1.0cm}
\includegraphics[height=.18\textheight, angle =0]{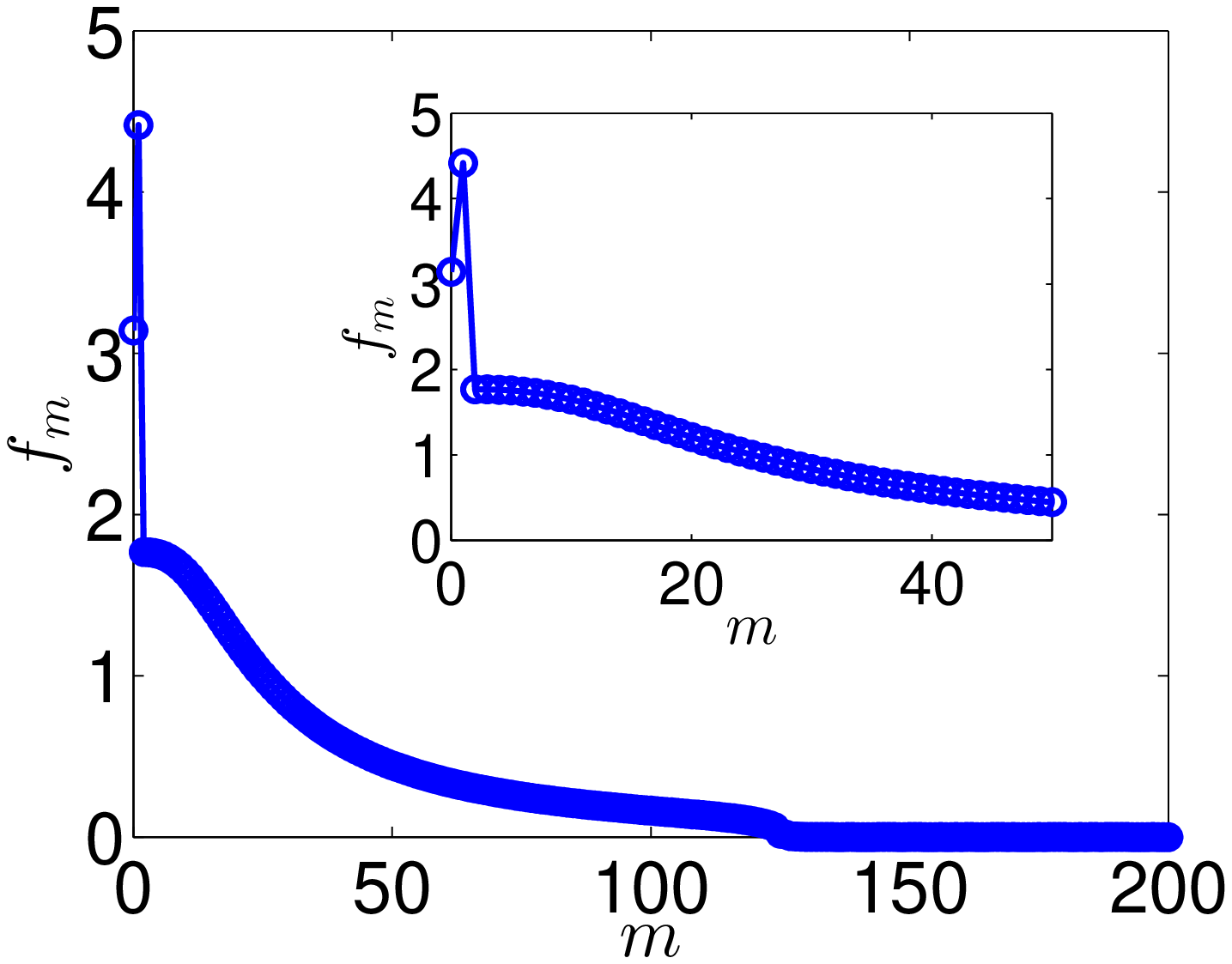}
}
\subfigure[][]{\hspace{-0.5cm}
\includegraphics[height=.18\textheight, angle =0]{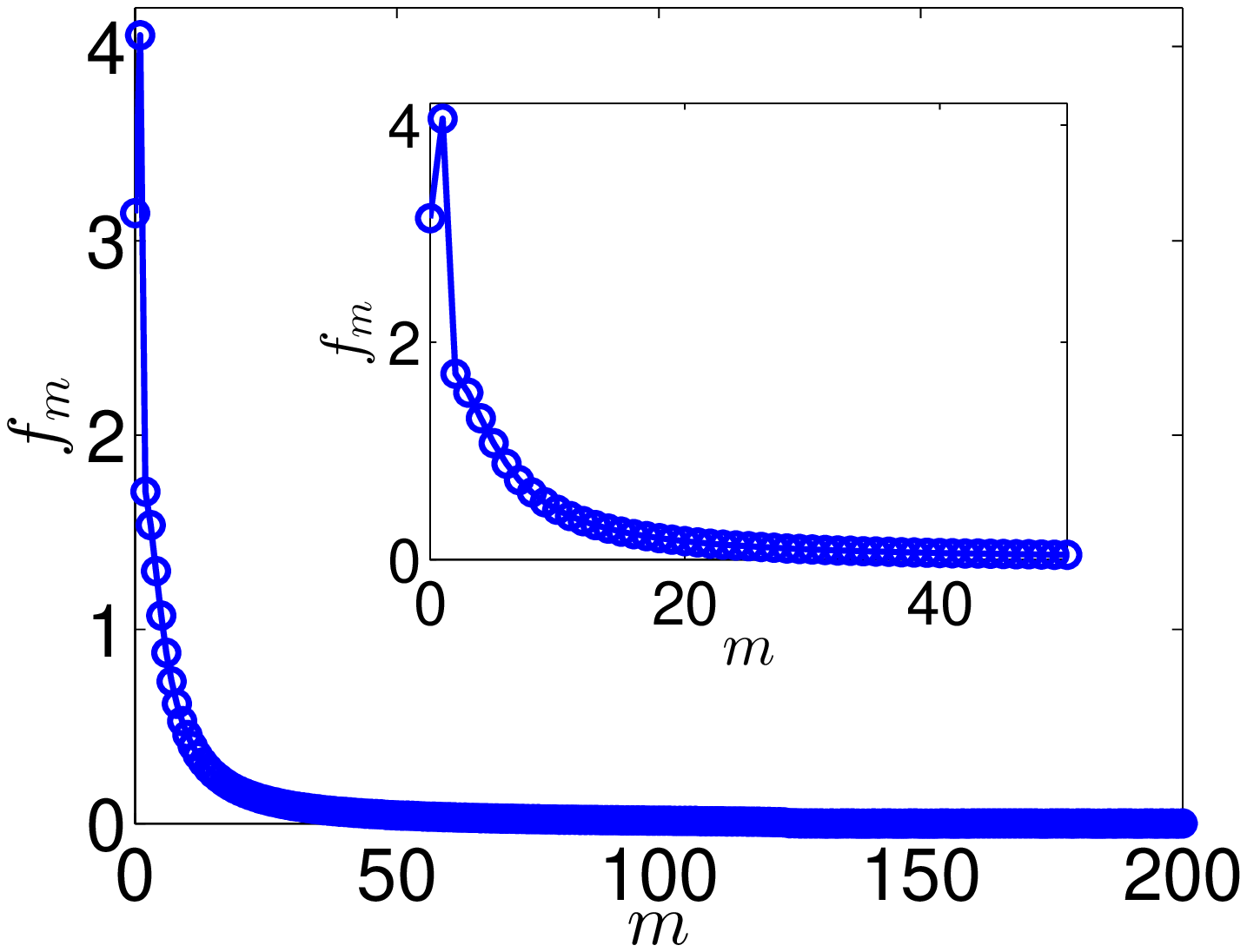}
}
\subfigure[][]{\hspace{-0.5cm}
\includegraphics[height=.18\textheight, angle =0]{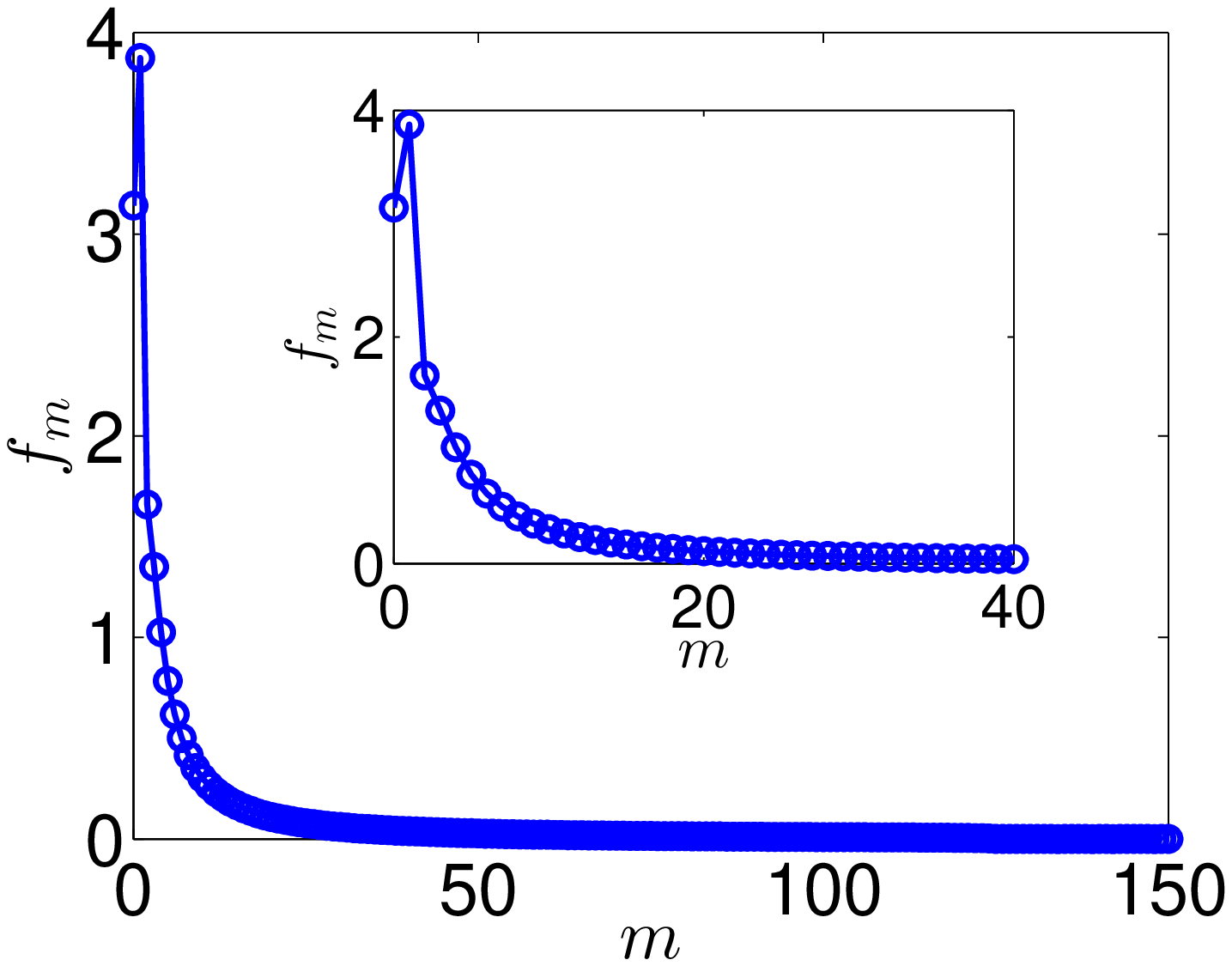}
}
}
\mbox{\hspace{0.0cm}
\subfigure[][]{\hspace{-1.0cm}
\includegraphics[height=.18\textheight, angle =0]{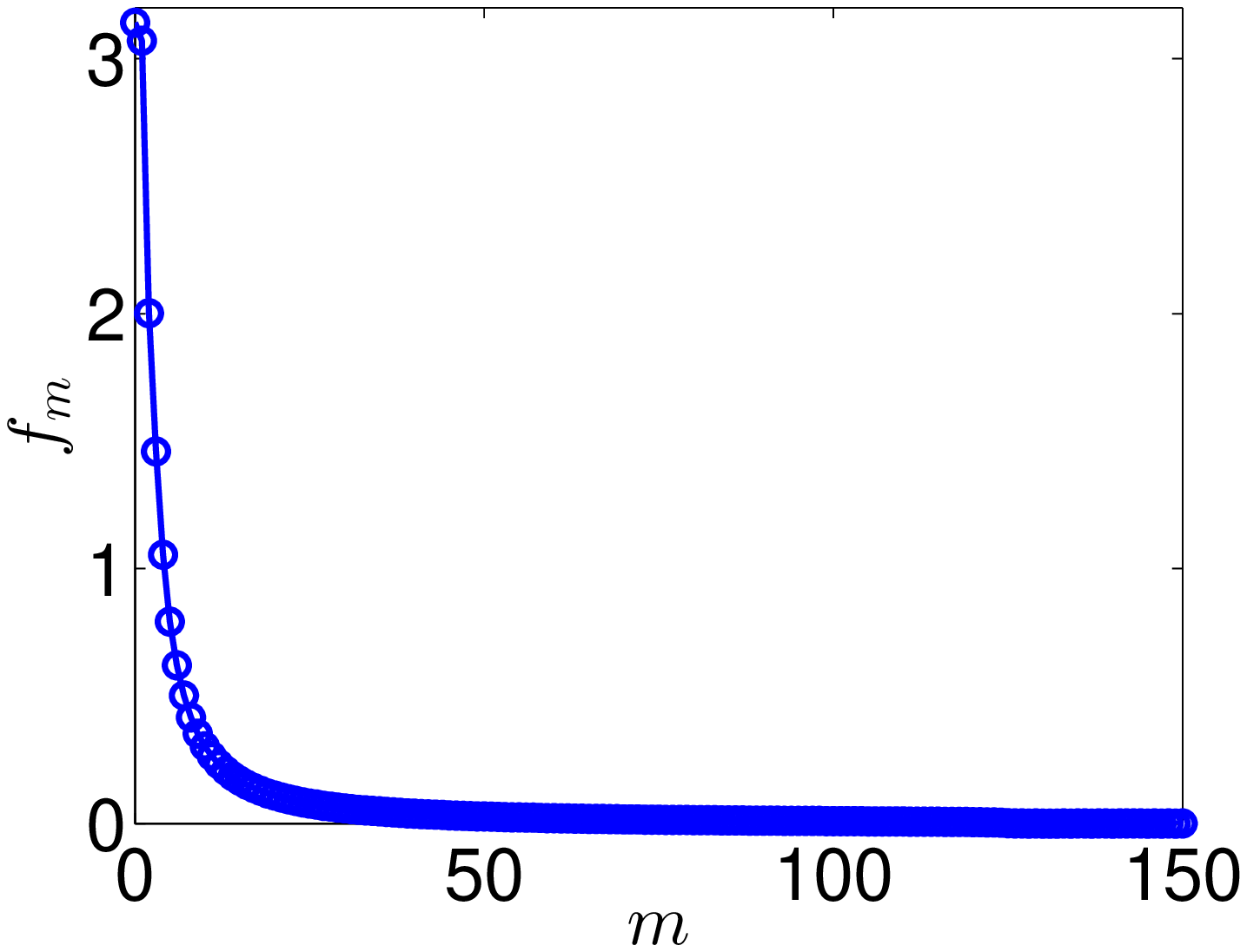}
}
\subfigure[][]{\hspace{-0.5cm}
\includegraphics[height=.18\textheight, angle =0]{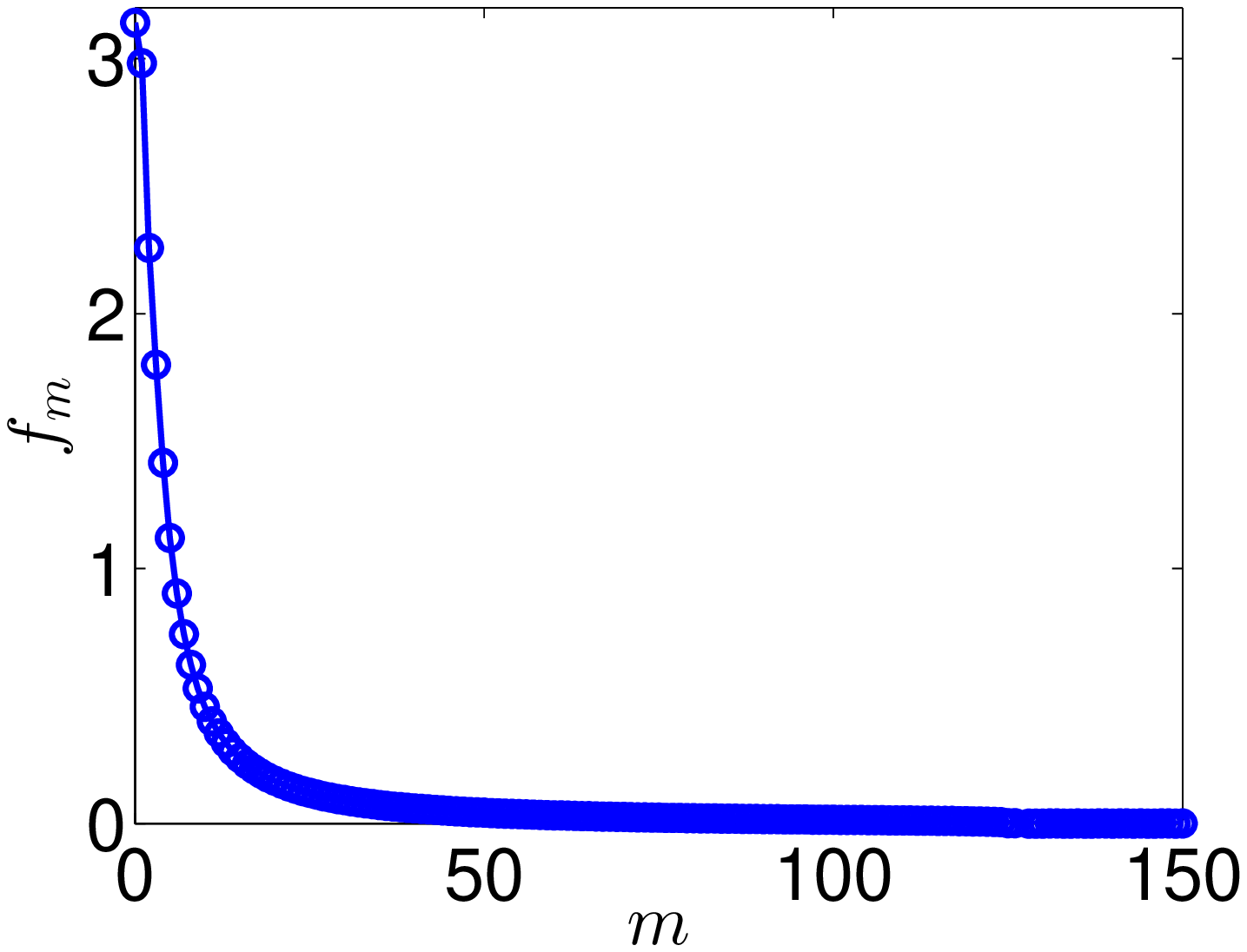}
}
\subfigure[][]{\hspace{-0.5cm}
\includegraphics[height=.18\textheight, angle =0]{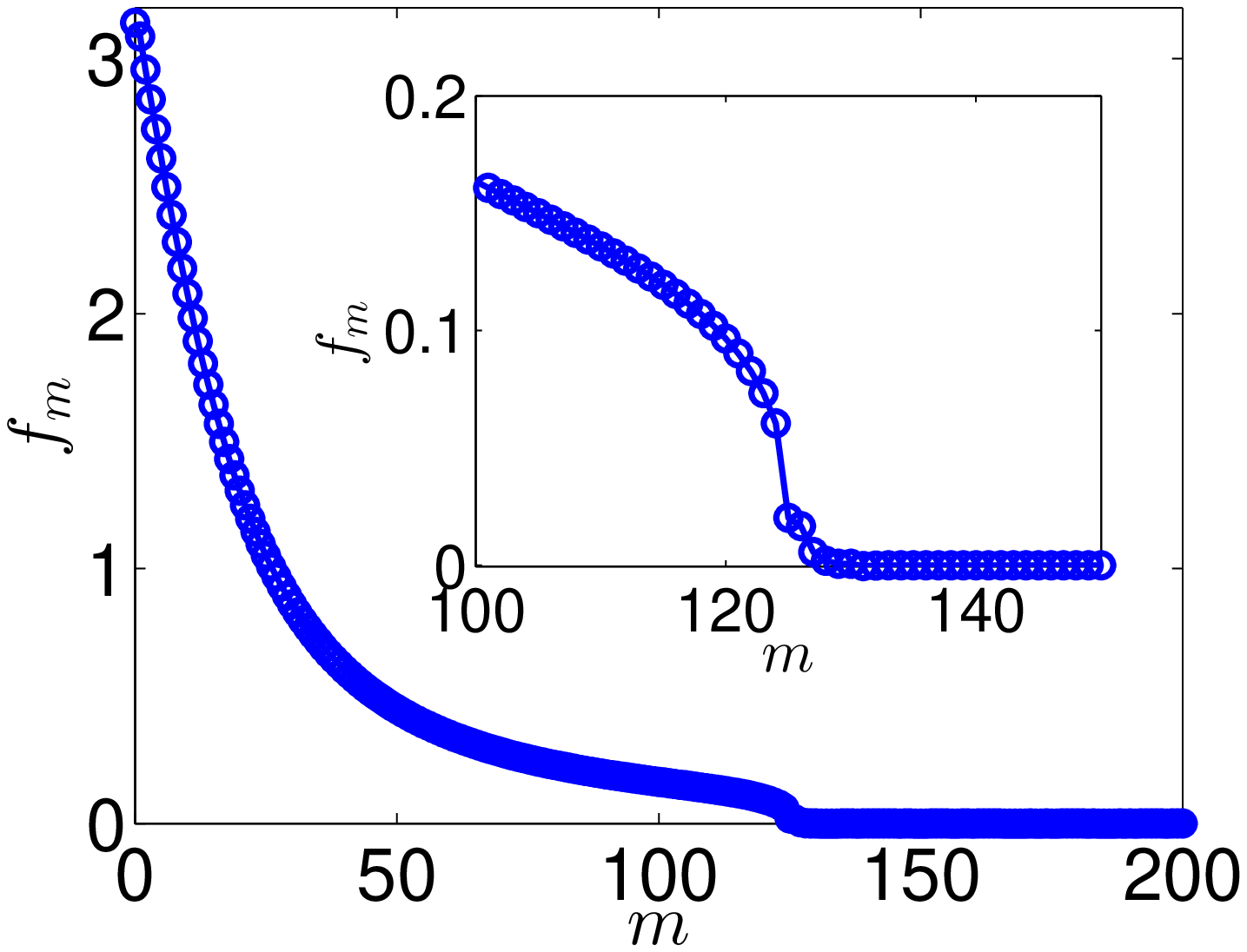}
}
}
\end{center}
\caption{
Same as Figure \ref{fig2} for  $\alpha=5$ branch (see, the right
panel of Figure \ref{fig1}). In particular, static discrete profiles 
for $h=0.06$: panels (a) and (f); $h=0.3$: panels (b) and (e); and  $h=0.4$:
panels (c) and (d).
\label{fig3}
}
\end{figure}

For instance, for the cases with $\alpha=4$ and $\alpha=5$, Figure~\ref{fig1}
showcases the functional dependence of $f_{1}$, i.e., the value of the profile
function at the first site in the domain, over the lattice spacing $h$ in the 
left and right panels, respectively. Furthermore, typical sample examples of static
lattice BPS skyrmions, i.e., plots of $f_{m}\doteq f(mh)$ against $m$, are shown in 
Figures~\ref{fig2} and \ref{fig3} for various
values of $h$ (see the captions in the relevant figures in connection with
the arrows appearing in Figure~\ref{fig1}).

These solutions provide a sense of the variation of the solution over the branch. 
In Figure \ref{fig2} (case of $\alpha=4$), the profile (a) is the most proximal to the continuum
limit, panels (b) and (c) arise in progressively more
discrete settings, while the ones of (d)-(f) give relevant
examples of the same solution branch past the fold point.
A similar phenomenology can be found in Figure \ref{fig3}
(case of $\alpha=5$), although now (f) represents the
solution most proximal to the continuum one. It is additionally
intriguing that a number of the solutions appear to have
a ``concave down'' profile as they tend to zero: that is shown in
panels (a) and (f), suggesting a
nearly compact waveform in the relevant solutions. 
In each case of $\alpha$, one of the branches
seems to be more ``coarse'' and more discrete in nature, while the
other is more proximal to the continuum limit.

\begin{figure}[!ht]
\begin{center}
\vspace{0.5cm}
\mbox{\hspace{0.0cm}
\subfigure[][]{\hspace{-0.5cm}
\includegraphics[height=.18\textheight, angle =0]{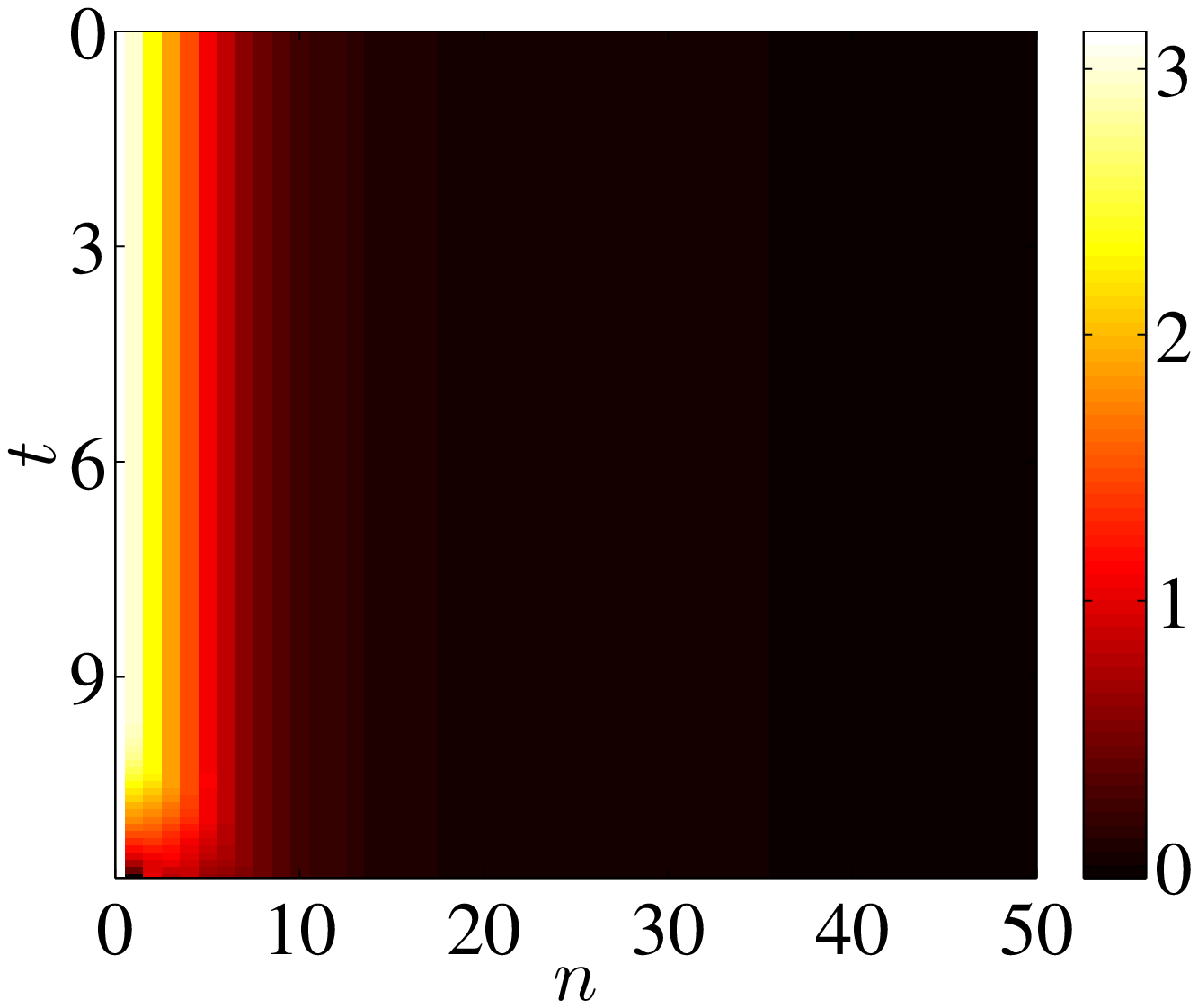}
}
\subfigure[][]{\hspace{-0.5cm}
\includegraphics[height=.18\textheight, angle =0]{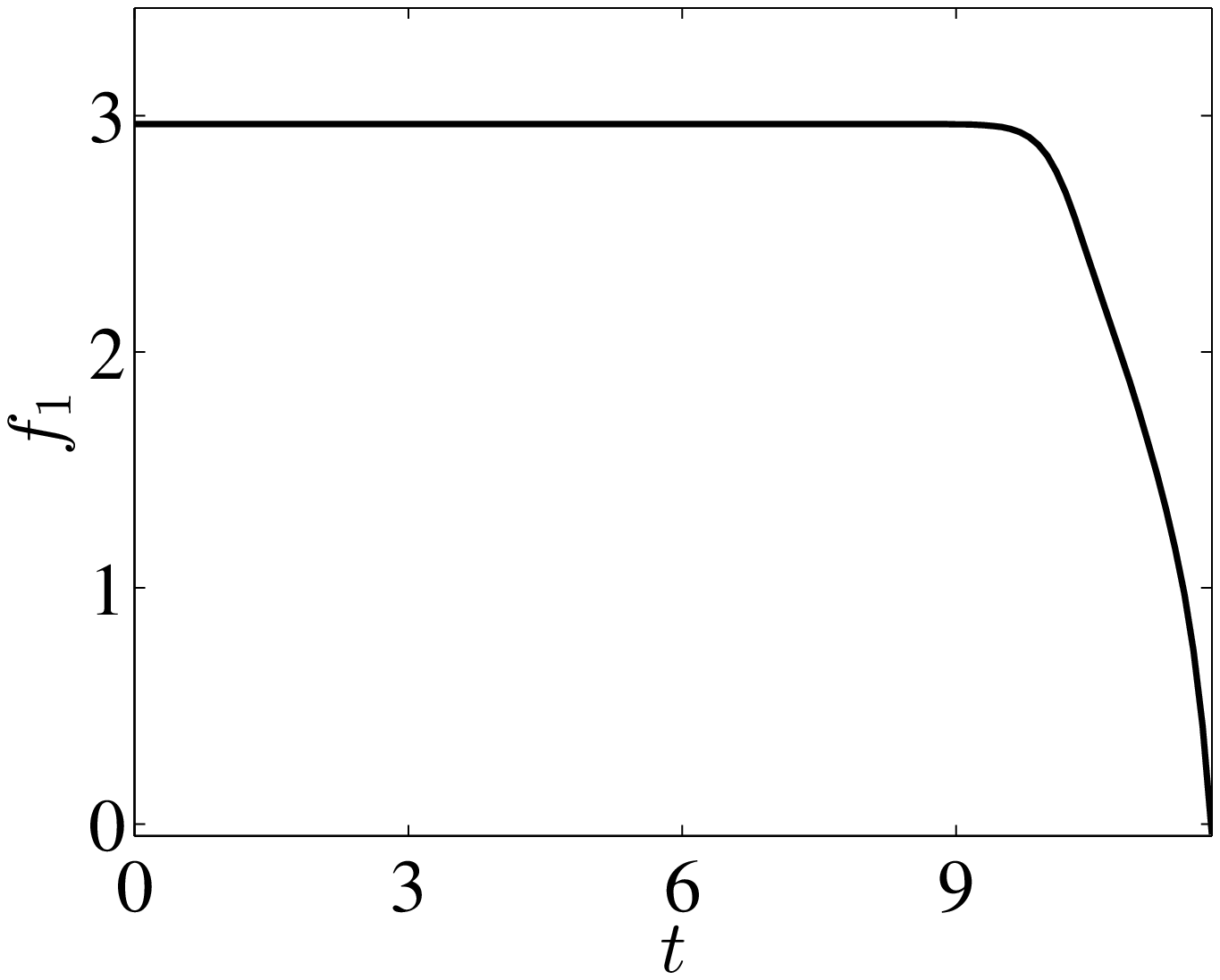}
}
\subfigure[][]{\hspace{-0.5cm}
\includegraphics[height=.18\textheight, angle =0]{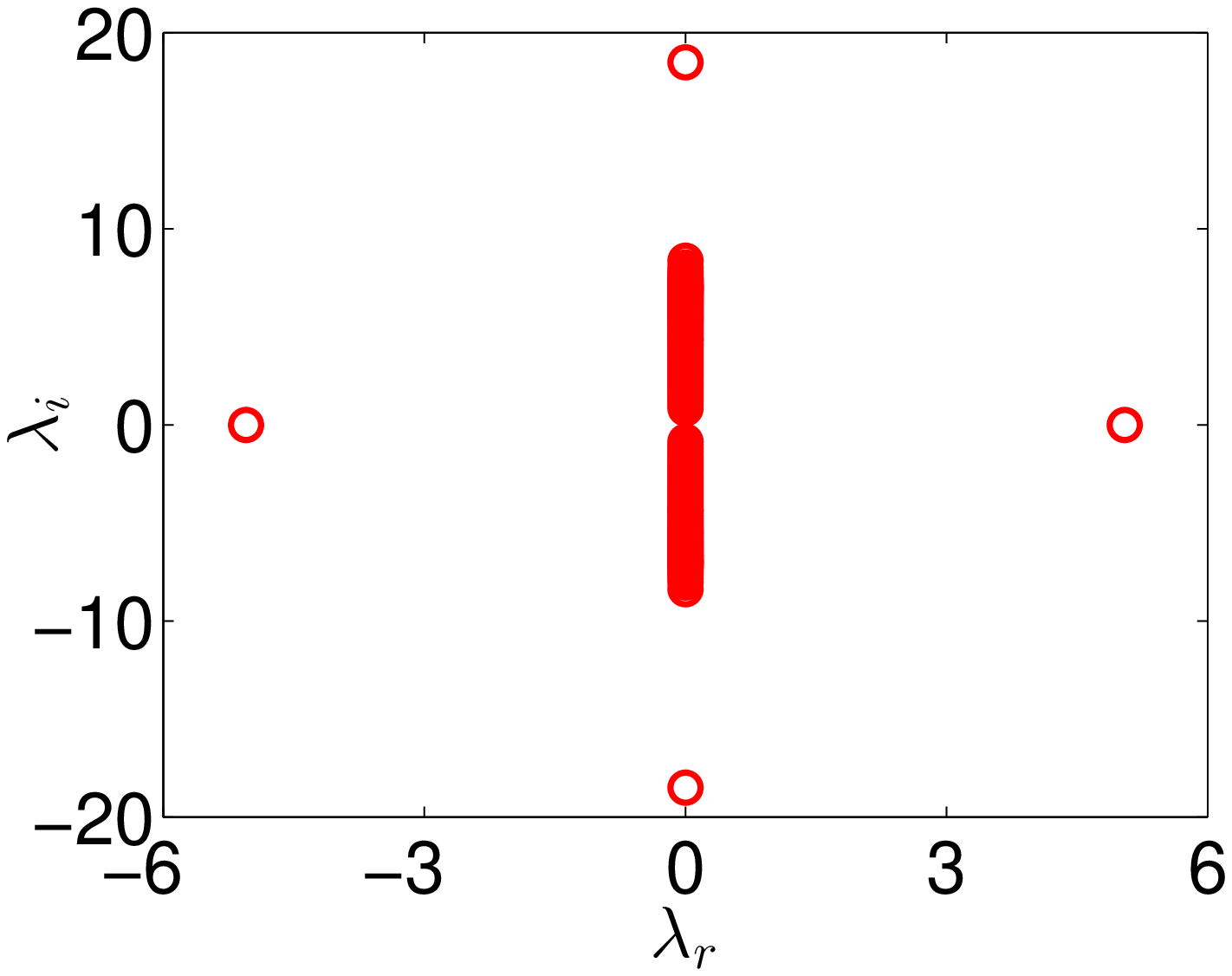}
}
}
\mbox{\hspace{0.0cm}
\subfigure[][]{\hspace{-0.5cm}
\includegraphics[height=.18\textheight, angle =0]{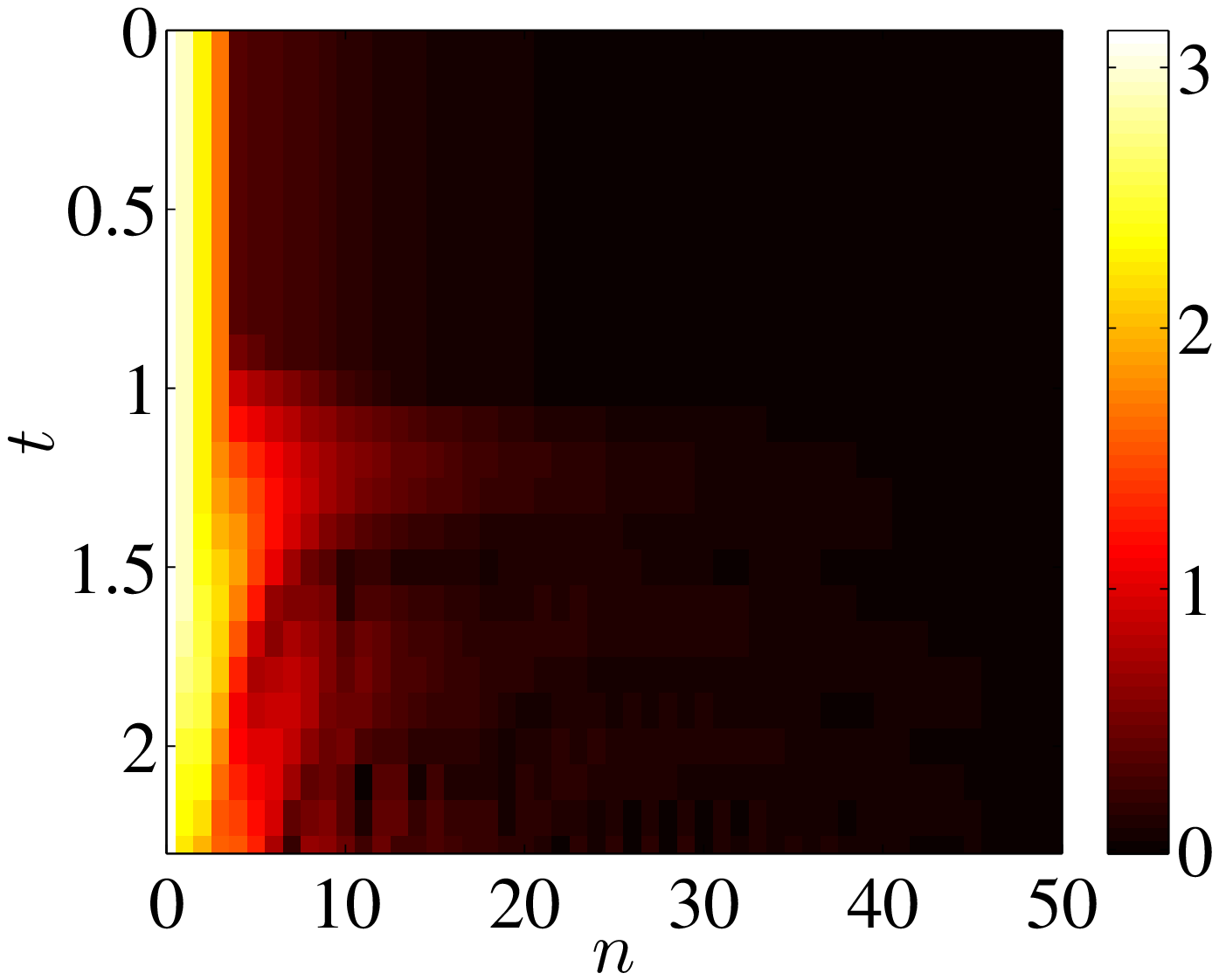}
}
\subfigure[][]{\hspace{-0.5cm}
\includegraphics[height=.18\textheight, angle =0]{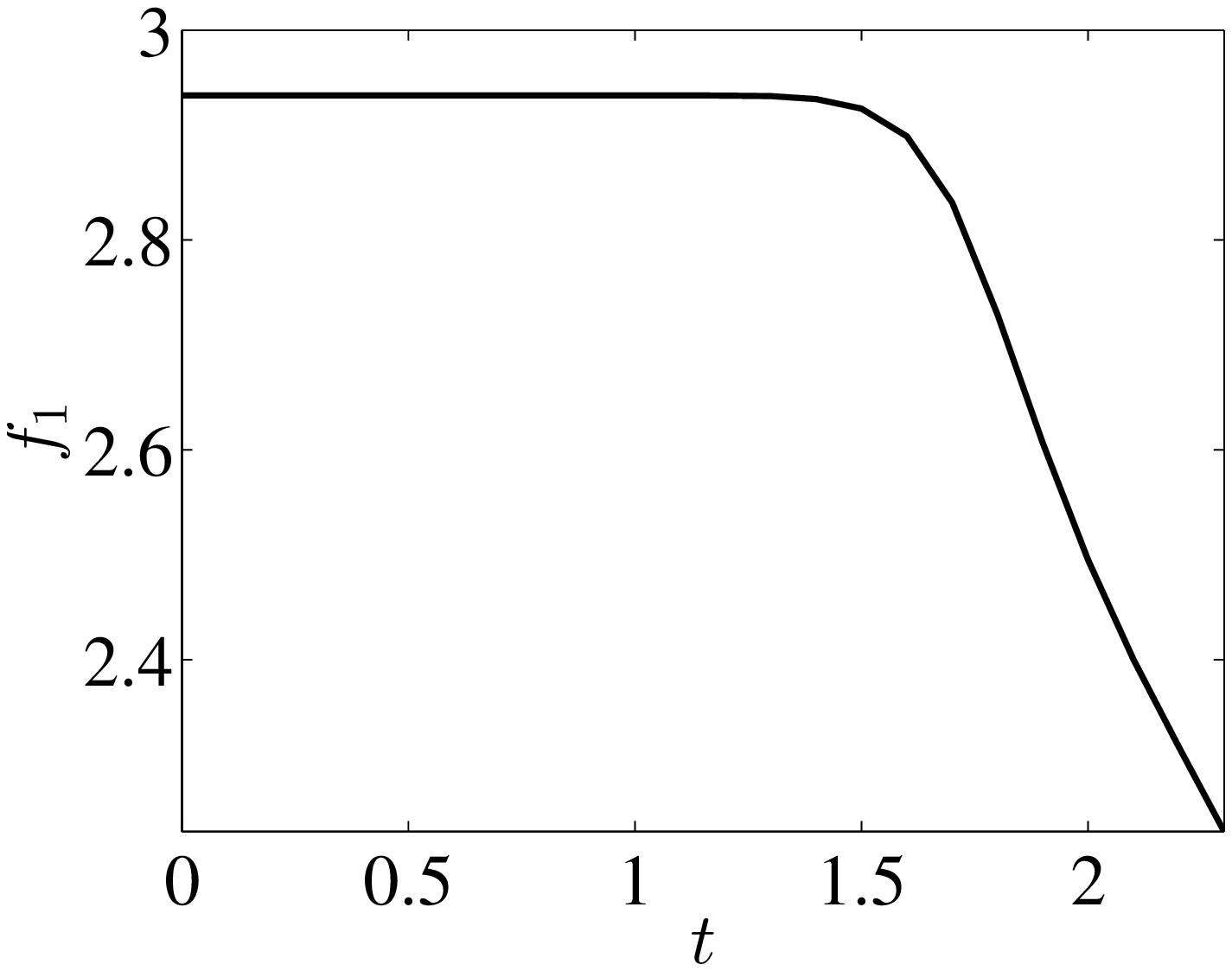}
}
\subfigure[][]{\hspace{-0.5cm}
\includegraphics[height=.18\textheight, angle =0]{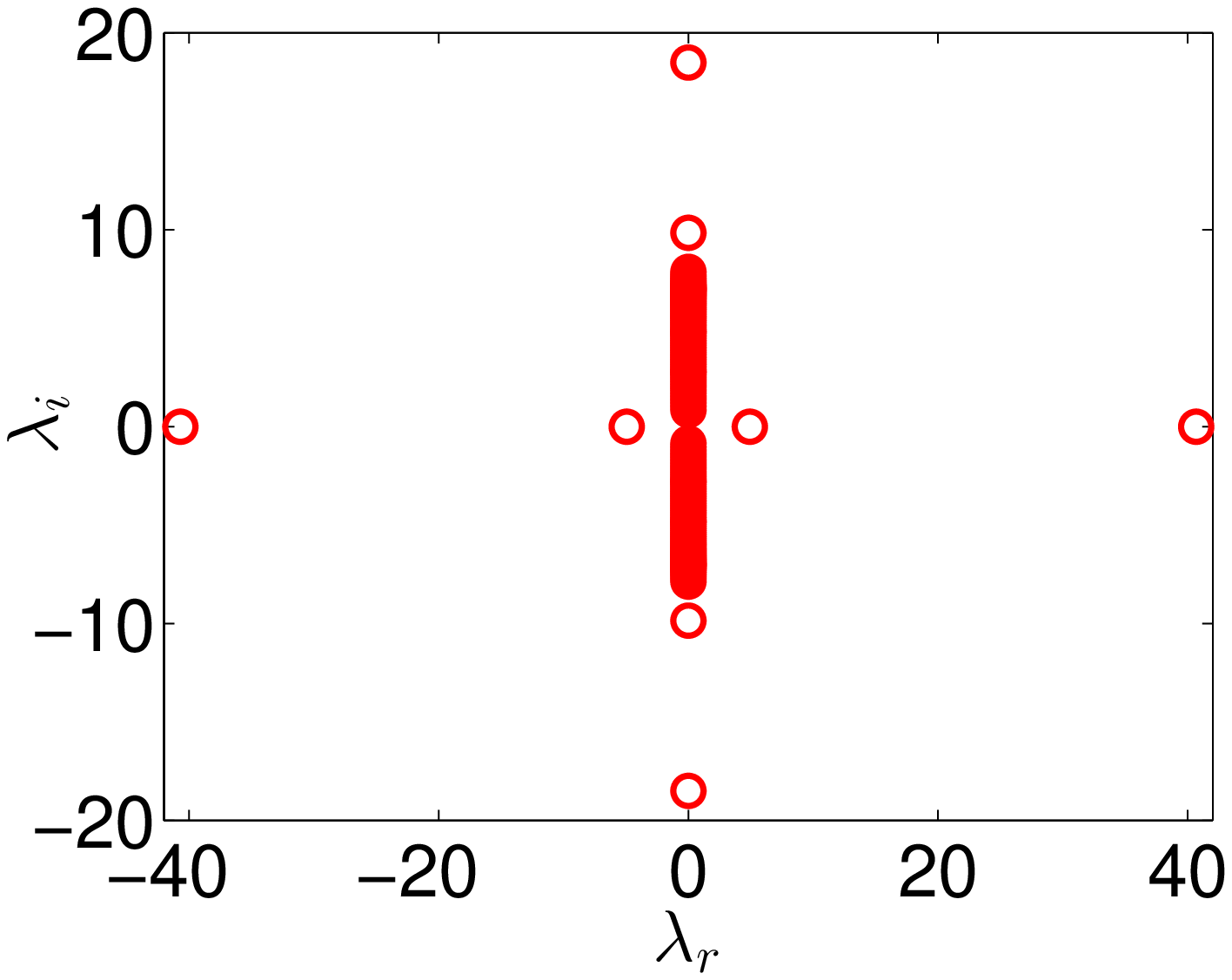}
}
}
\end{center}
\caption{
Spatio-temporal evolution of the steady-state solutions of panels (b)
and (e) of Figure \ref{fig2} and associated spectra for $\alpha=4$ and
$h=0.3$ are shown at the top and bottom panels, respectively. The left
and middle panels correspond to the space-time evolution of the profile 
function and the dependence of $f_{1}$ on time $t$, respectively, 
whereas the right panels present the corresponding spectra.
\label{fig4}
}
\end{figure}

\begin{figure}[!t]
\begin{center}
\vspace{0.5cm}
\mbox{\hspace{0.0cm}
\subfigure[][]{\hspace{-0.5cm}
\includegraphics[height=.18\textheight, angle =0]{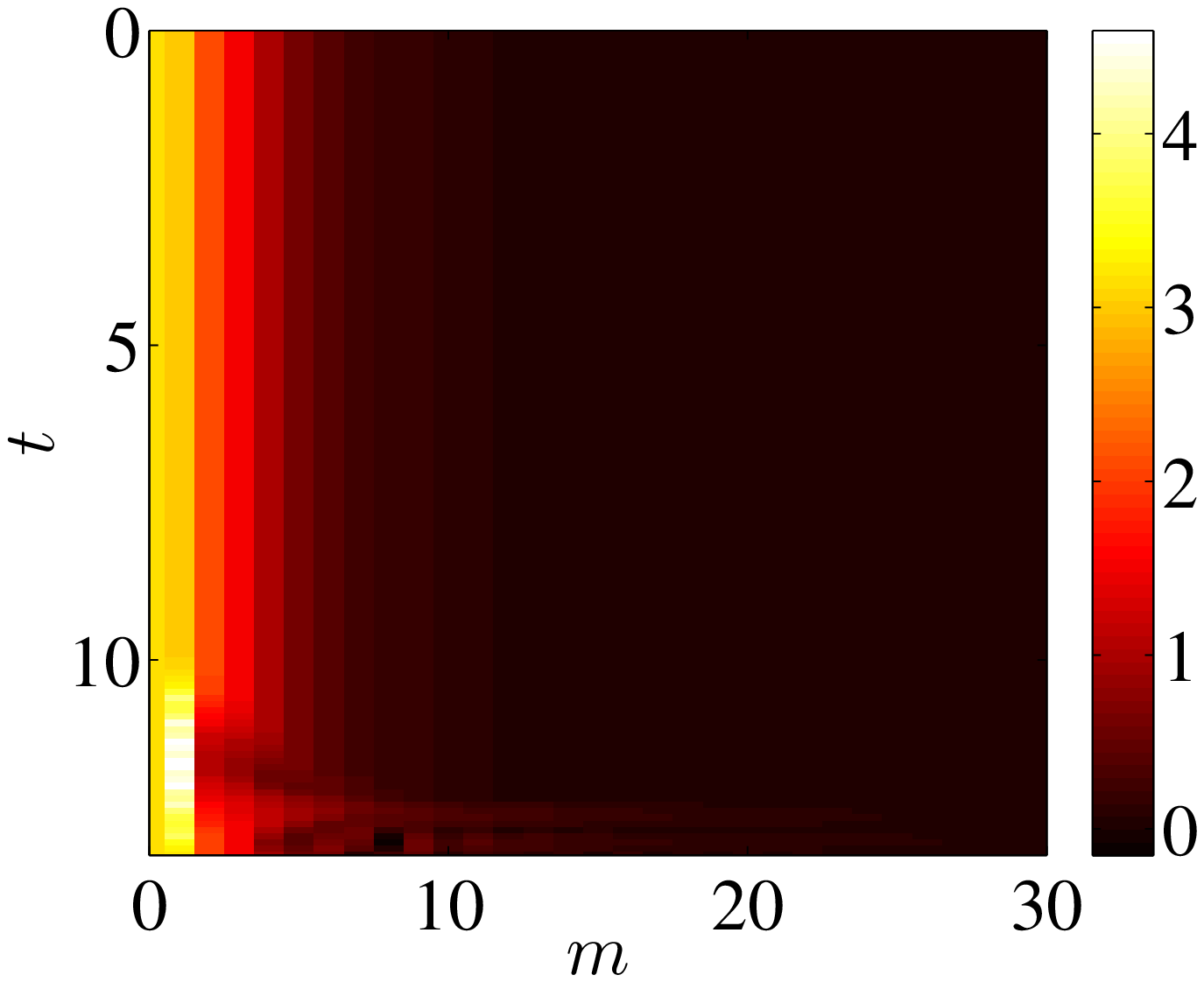}
}
\subfigure[][]{\hspace{-0.5cm}
\includegraphics[height=.18\textheight, angle =0]{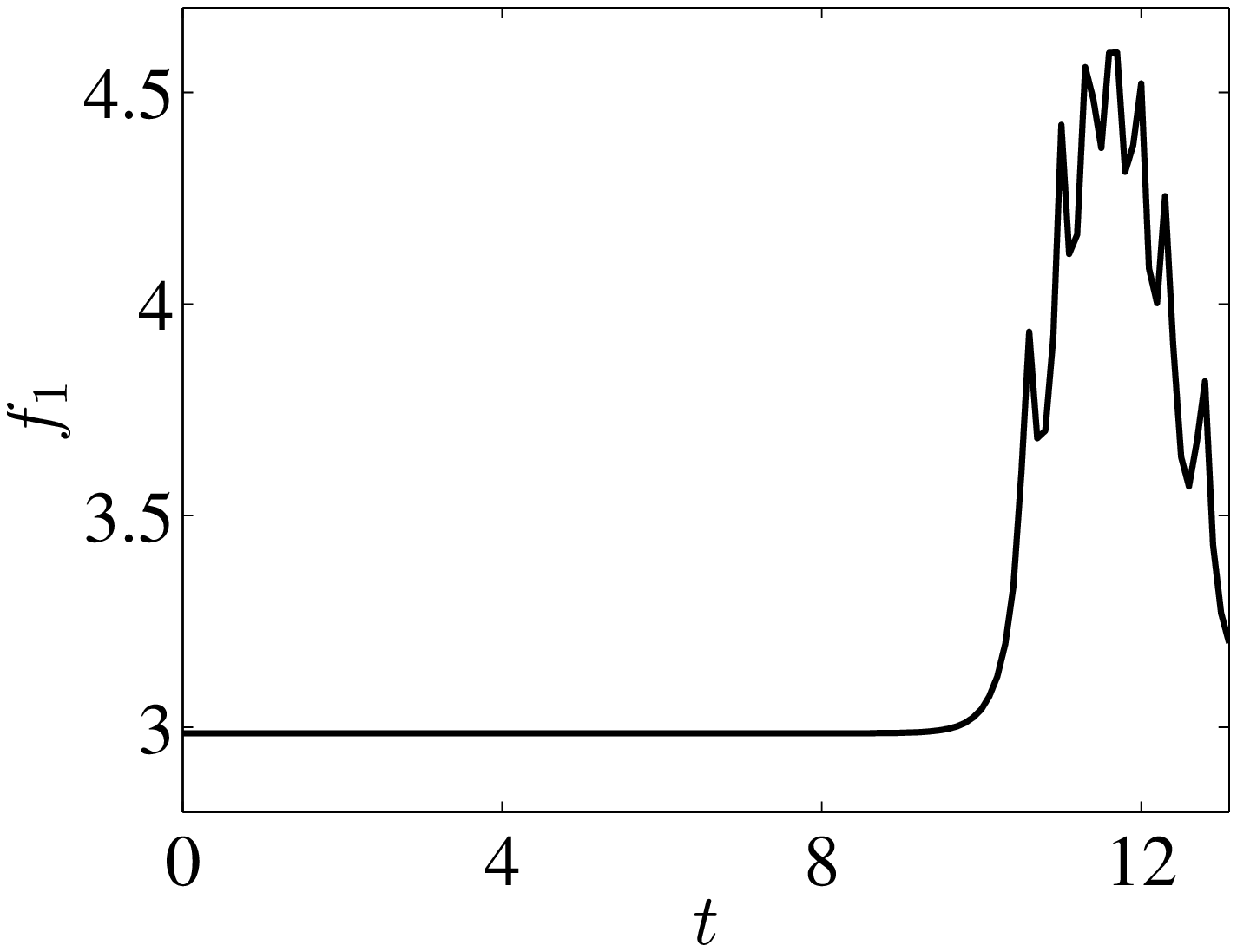}
}
\subfigure[][]{\hspace{-0.5cm}
\includegraphics[height=.18\textheight, angle =0]{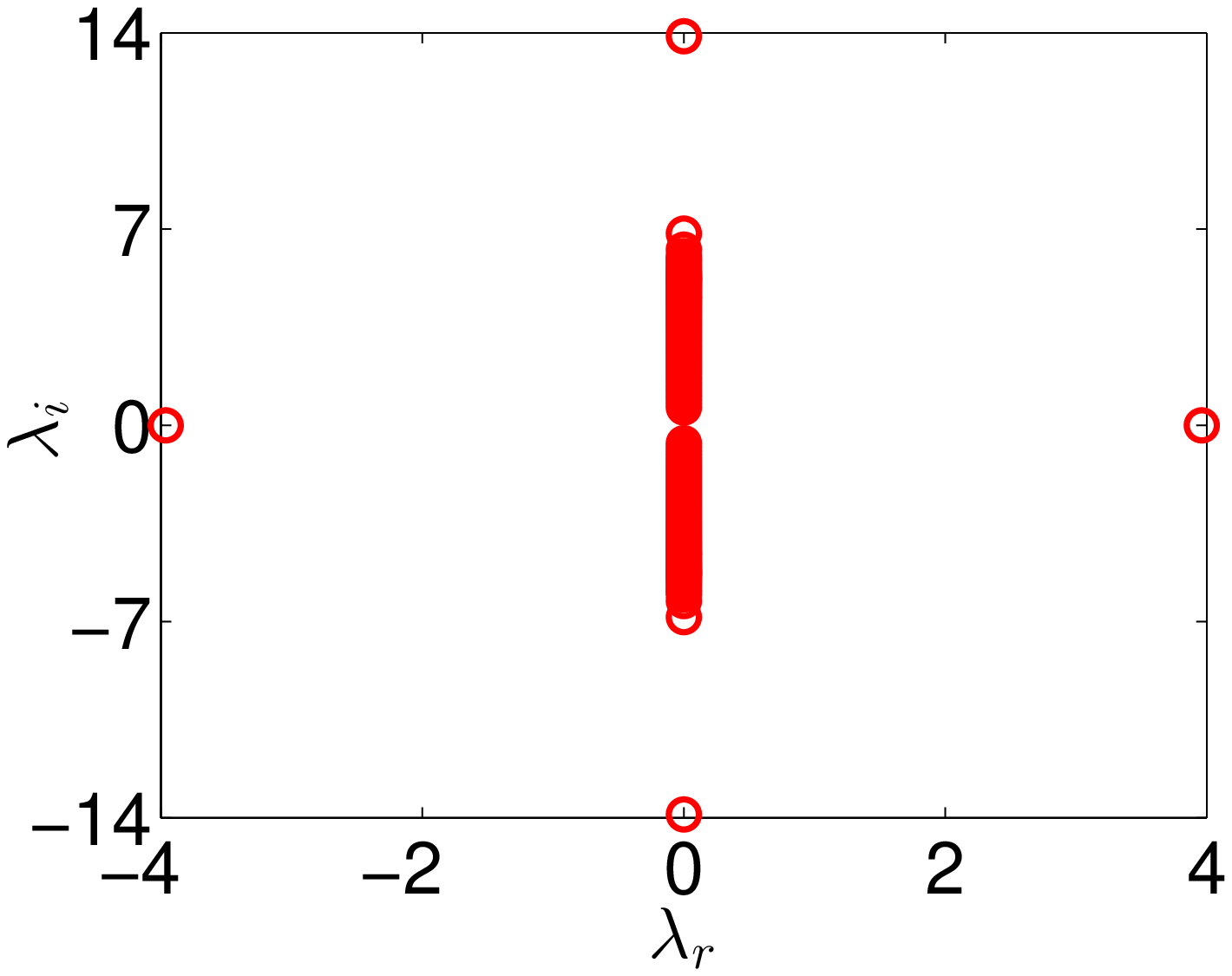}
}
}
\mbox{\hspace{0.0cm}
\subfigure[][]{\hspace{-0.5cm}
\includegraphics[height=.18\textheight, angle =0]{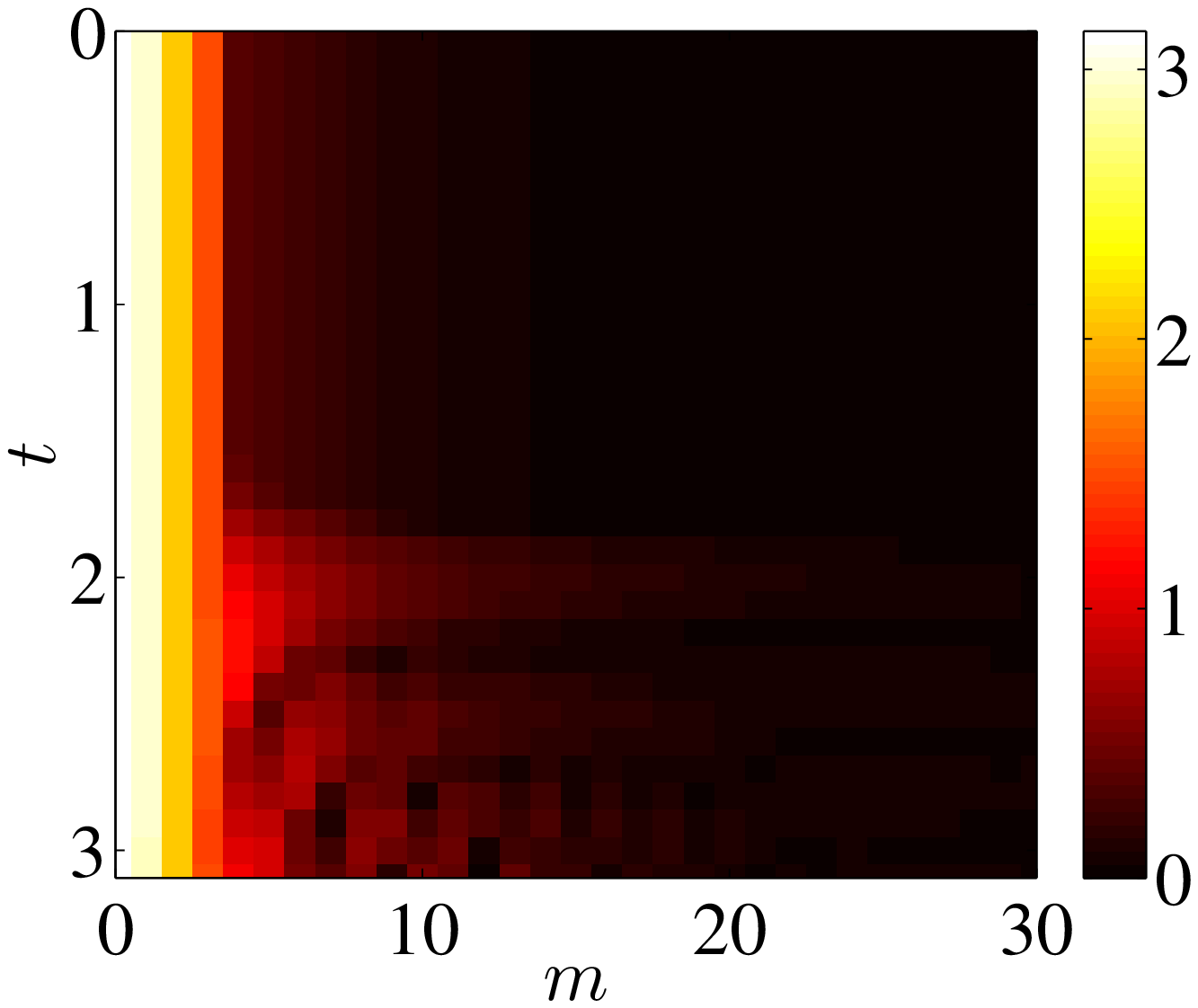}
}
\subfigure[][]{\hspace{-0.5cm}
\includegraphics[height=.18\textheight, angle =0]{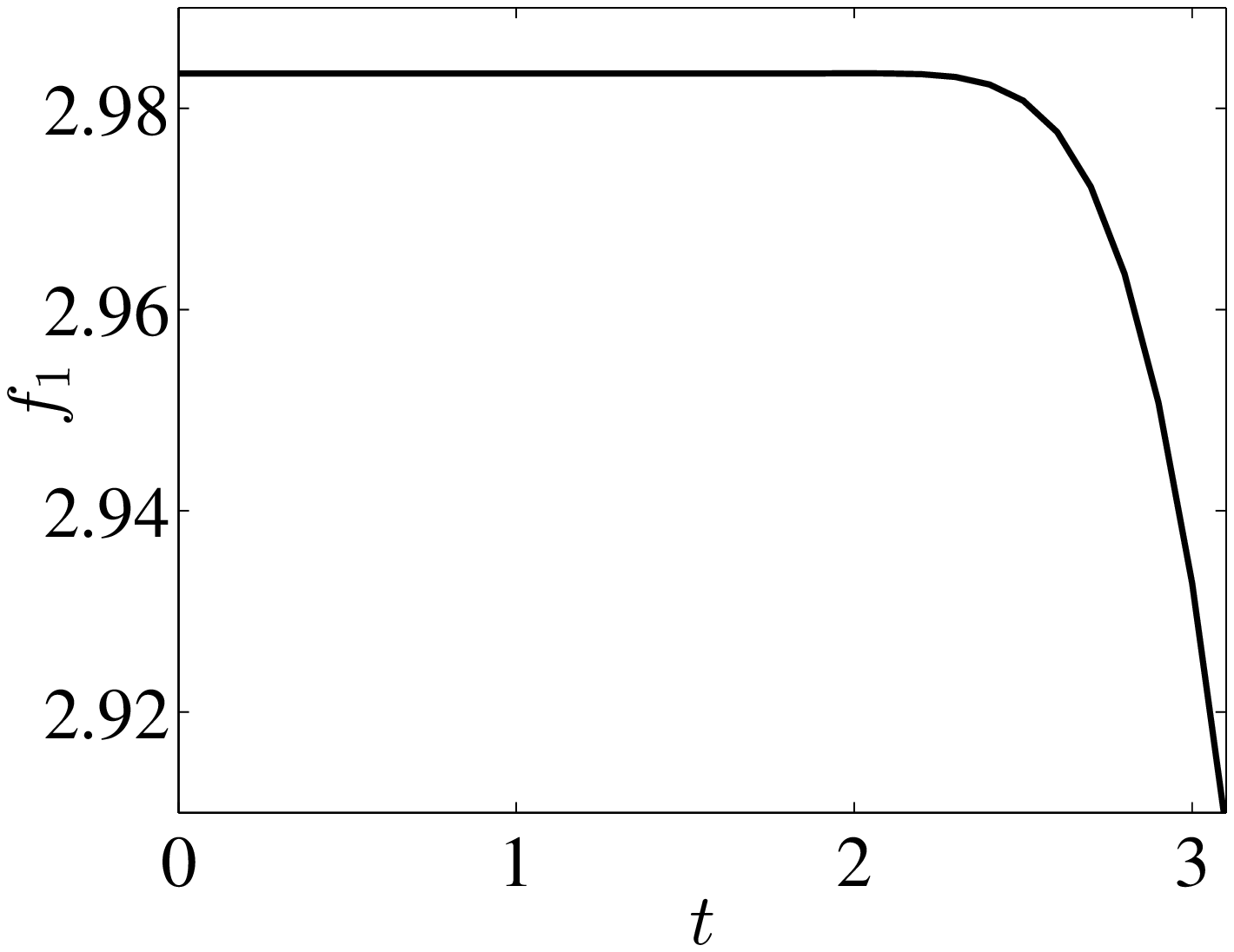}
}
\subfigure[][]{\hspace{-0.5cm}
\includegraphics[height=.18\textheight, angle =0]{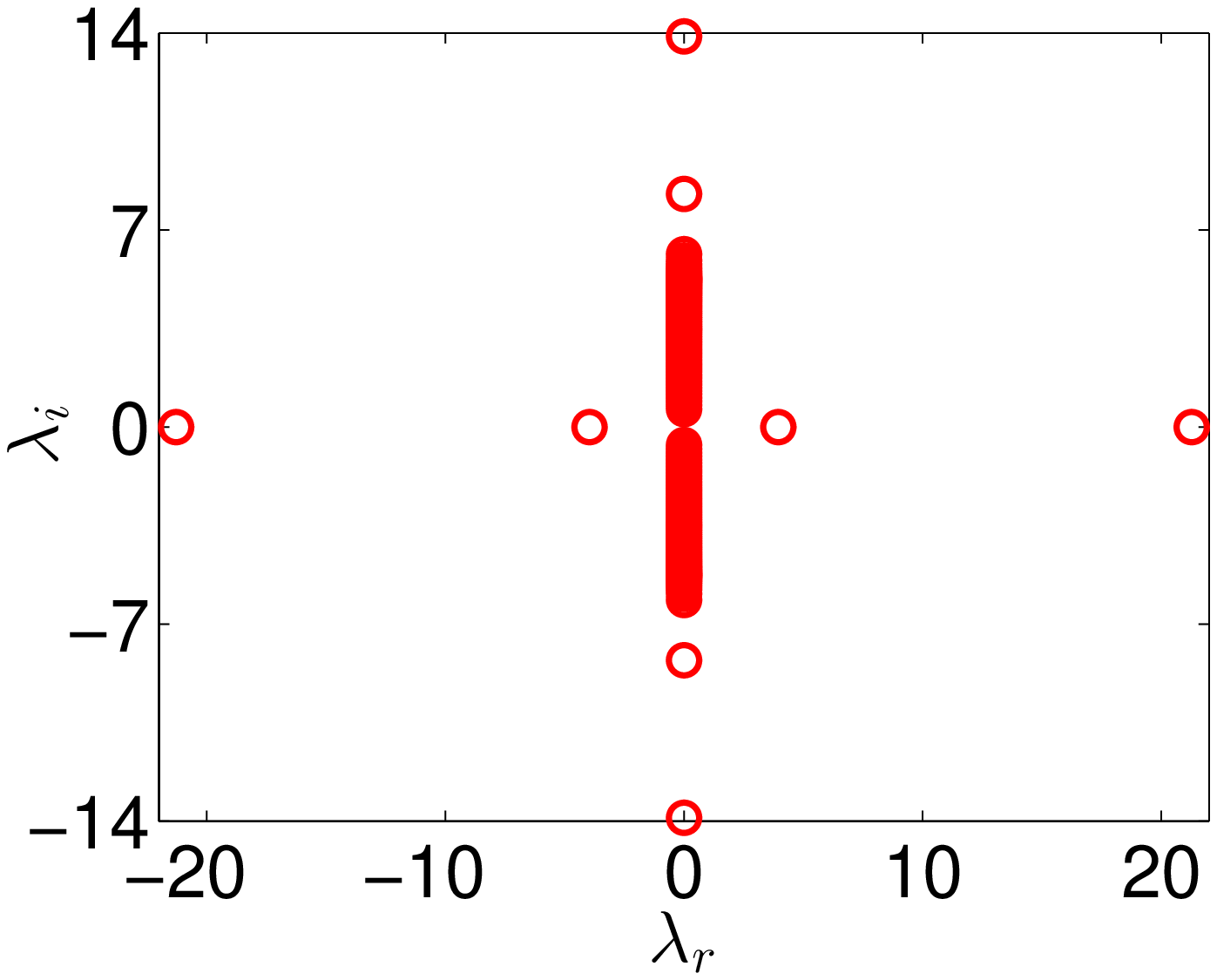}
}
}
\end{center}
\caption{
Same as Fig.~\ref{fig4} but for $h=0.4$. Spatiotemporal evolution of 
the steady-state solutions of panels (c) and (d) of Figure~\ref{fig2}
and associated spectra are shown at the top and bottom panels, respectively. 
The left and middle panels correspond to the space-time evolution of the 
profile function and the dependence of $f_{1}$ on time $t$, respectively,
whereas the right panels present the underlying spectra.
\label{fig5}
}
\end{figure}

\begin{figure}[!t]
\begin{center}
\vspace{0.5cm}
\mbox{\hspace{0.0cm}
\subfigure[][]{\hspace{-0.5cm}
\includegraphics[height=.18\textheight, angle =0]{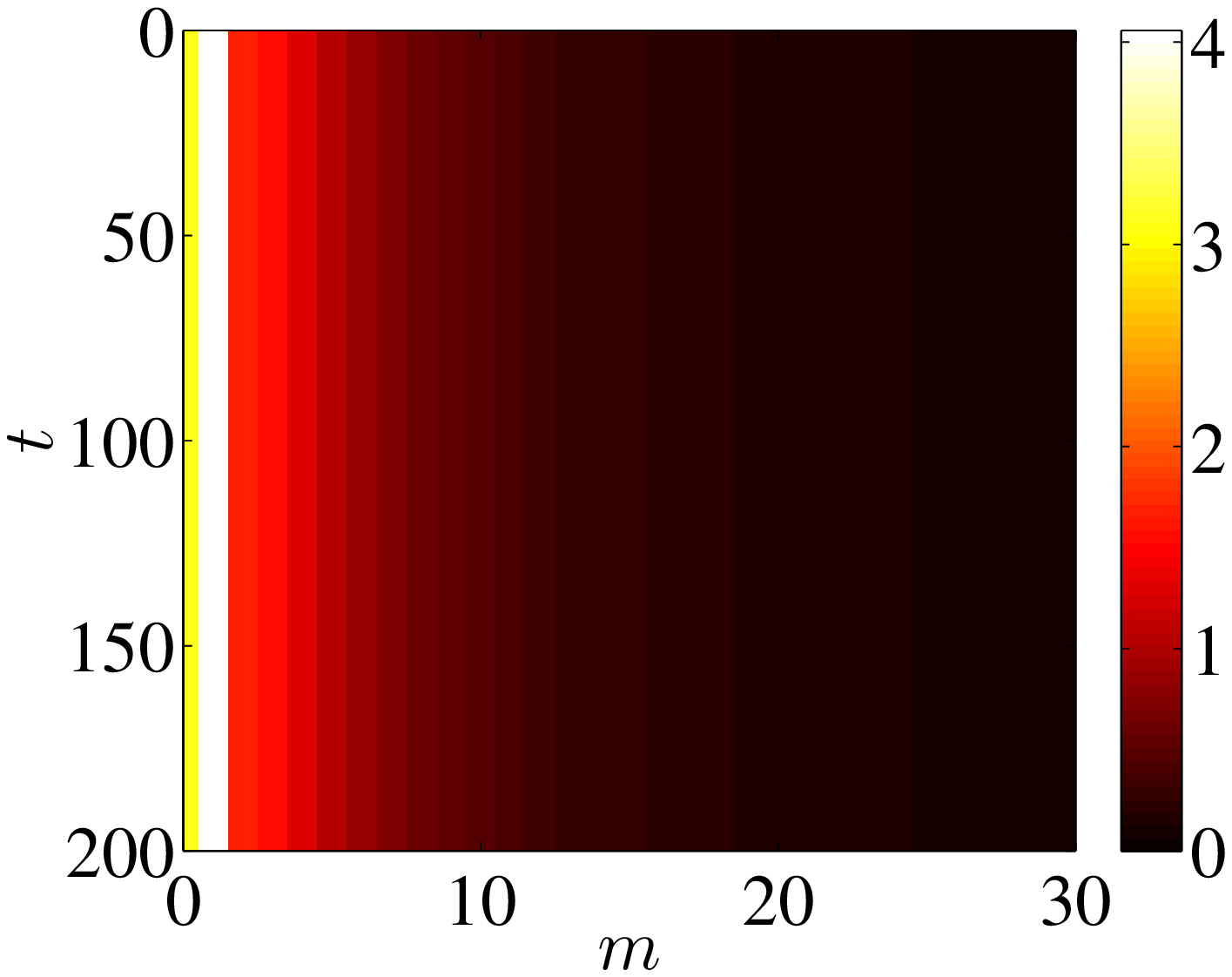}
}
\subfigure[][]{\hspace{-0.5cm}
\includegraphics[height=.18\textheight, angle =0]{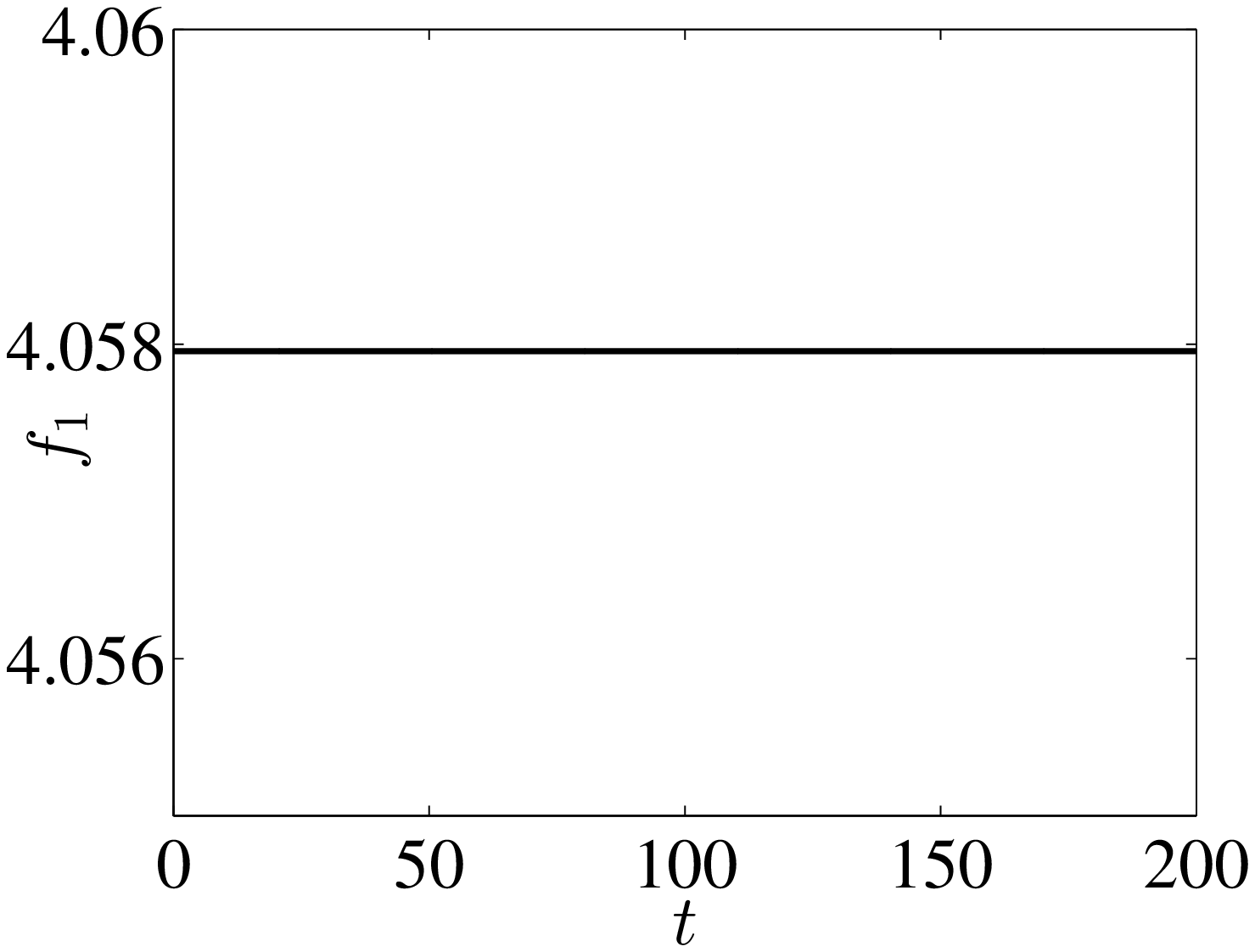}
}
\subfigure[][]{\hspace{-0.5cm}
\includegraphics[height=.18\textheight, angle =0]{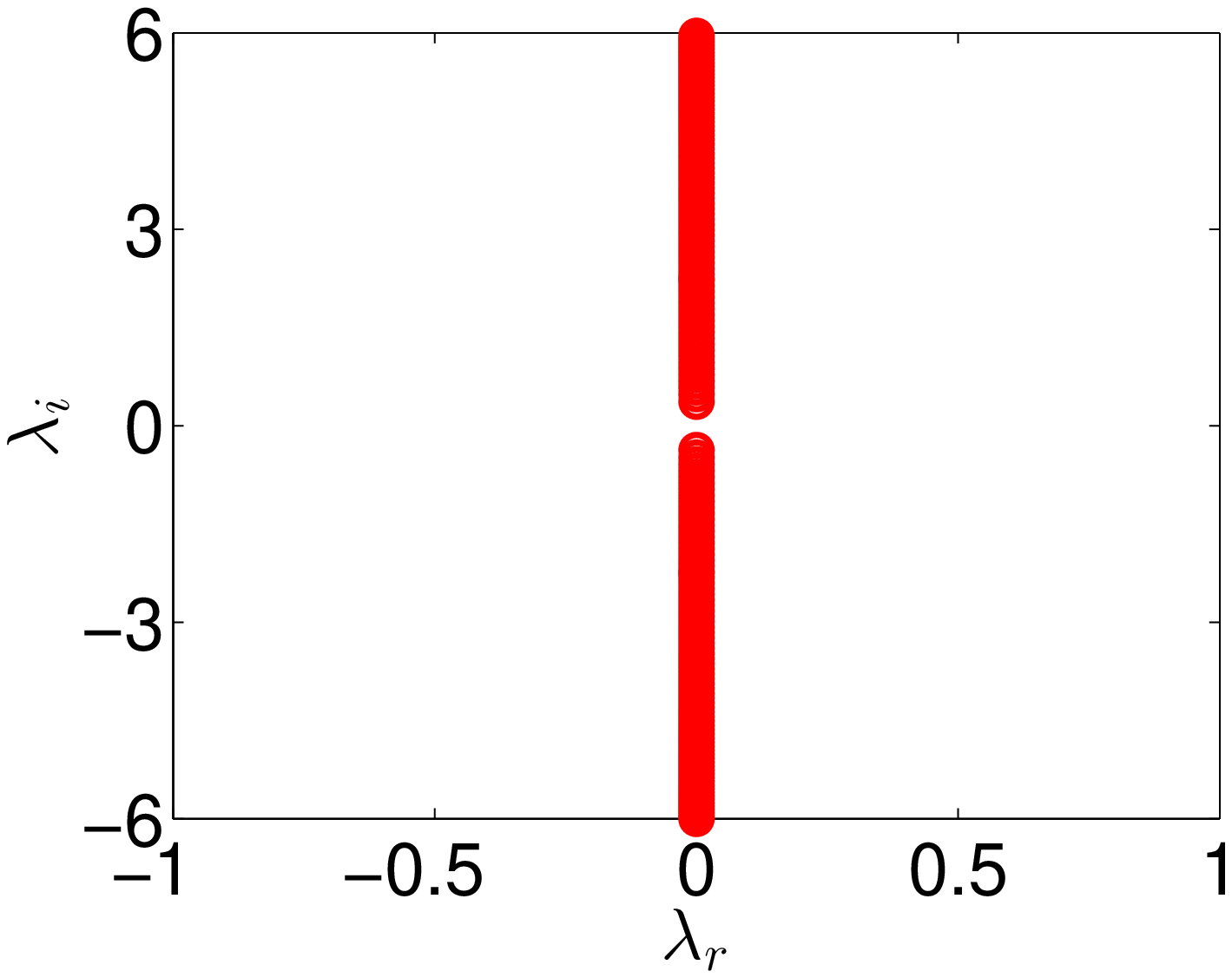}
}
}
\mbox{\hspace{0.0cm}
\subfigure[][]{\hspace{-0.5cm}
\includegraphics[height=.18\textheight, angle =0]{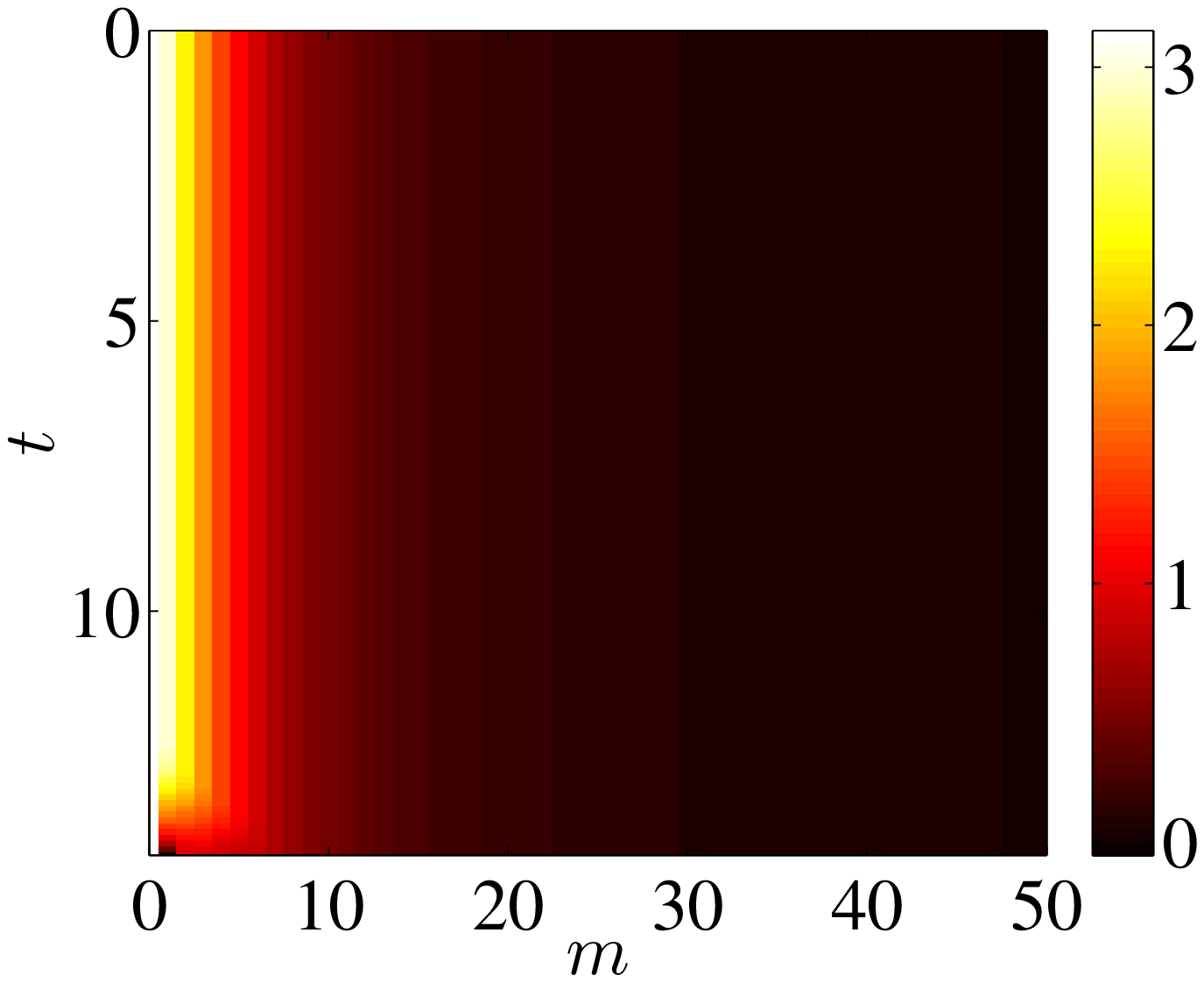}
}
\subfigure[][]{\hspace{-0.5cm}
\includegraphics[height=.18\textheight, angle =0]{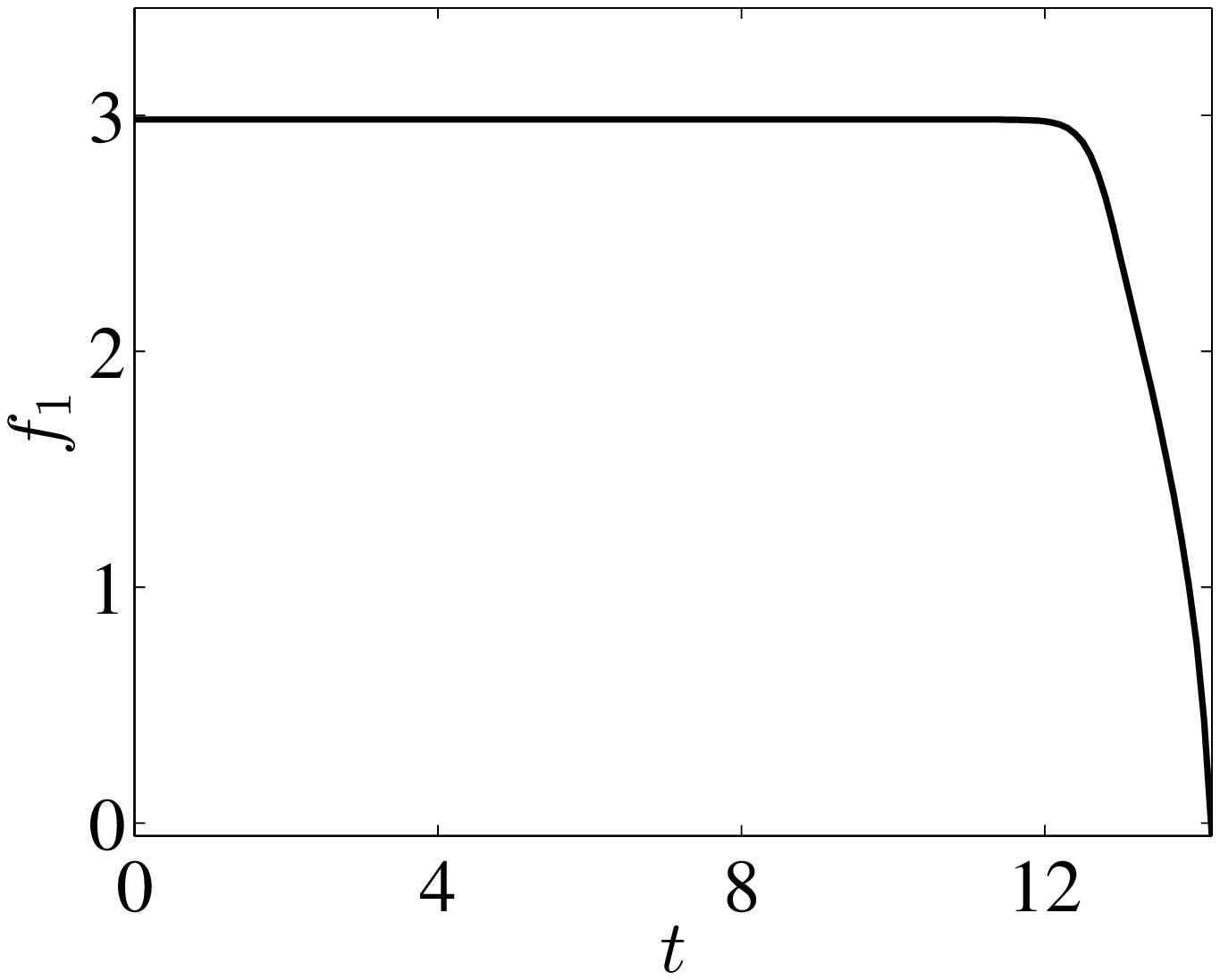}
}
\subfigure[][]{\hspace{-0.5cm}
\includegraphics[height=.18\textheight, angle =0]{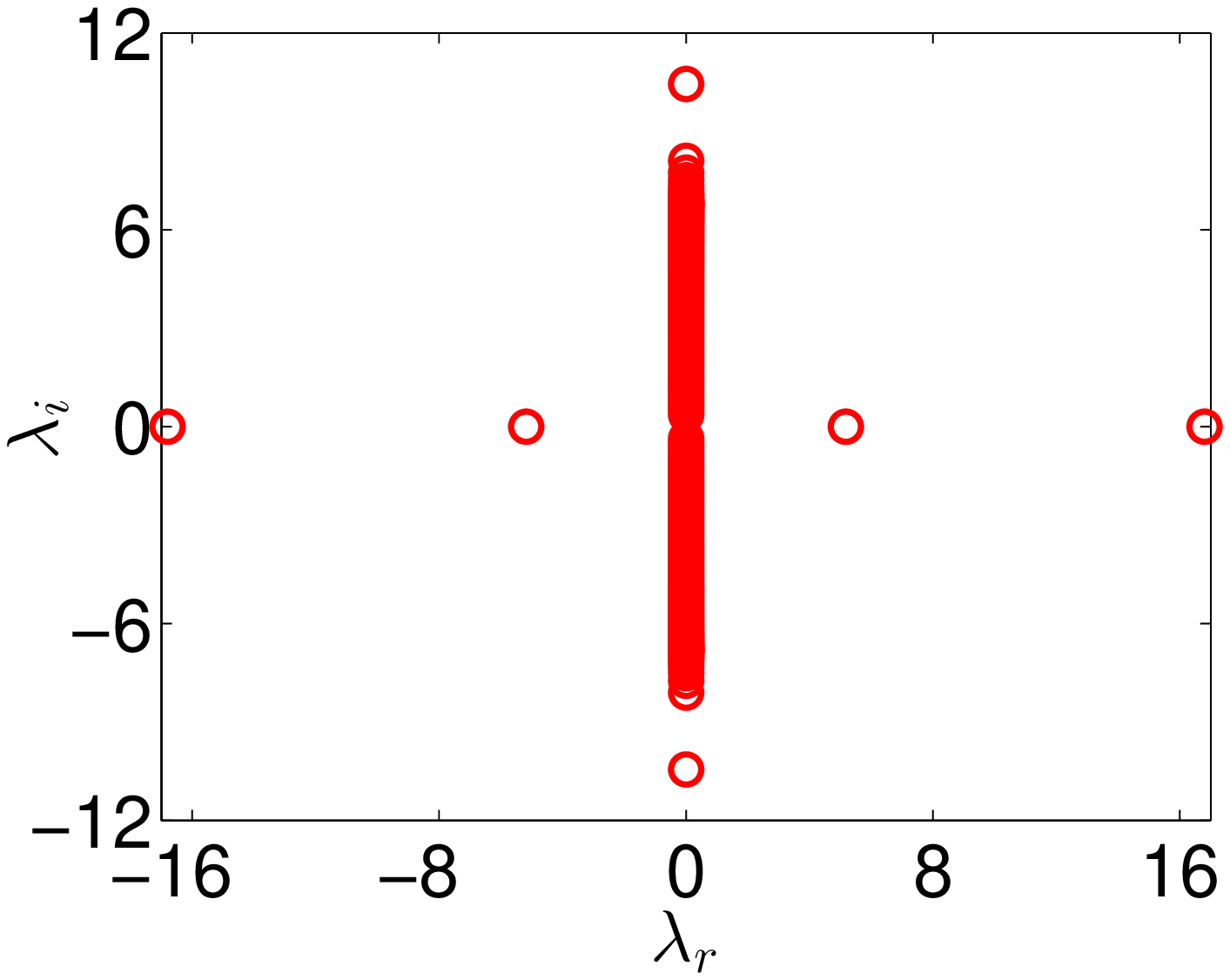}
}
}
\end{center}
\caption{
Same as Fig.~\ref{fig4} but for $\alpha=5$ and $h=0.3$. Spatio-temporal 
evolution of the steady-state solutions of panels (b) and (e) of Figure~\ref{fig3}
and associated spectra are shown at the top and bottom panels, respectively. 
Left, middle and right panels are the same as in the previous figures.
\label{fig6}
}
\end{figure}

\begin{figure}[!t]
\begin{center}
\vspace{0.5cm}
\mbox{\hspace{0.0cm}
\subfigure[][]{\hspace{-0.5cm}
\includegraphics[height=.18\textheight, angle =0]{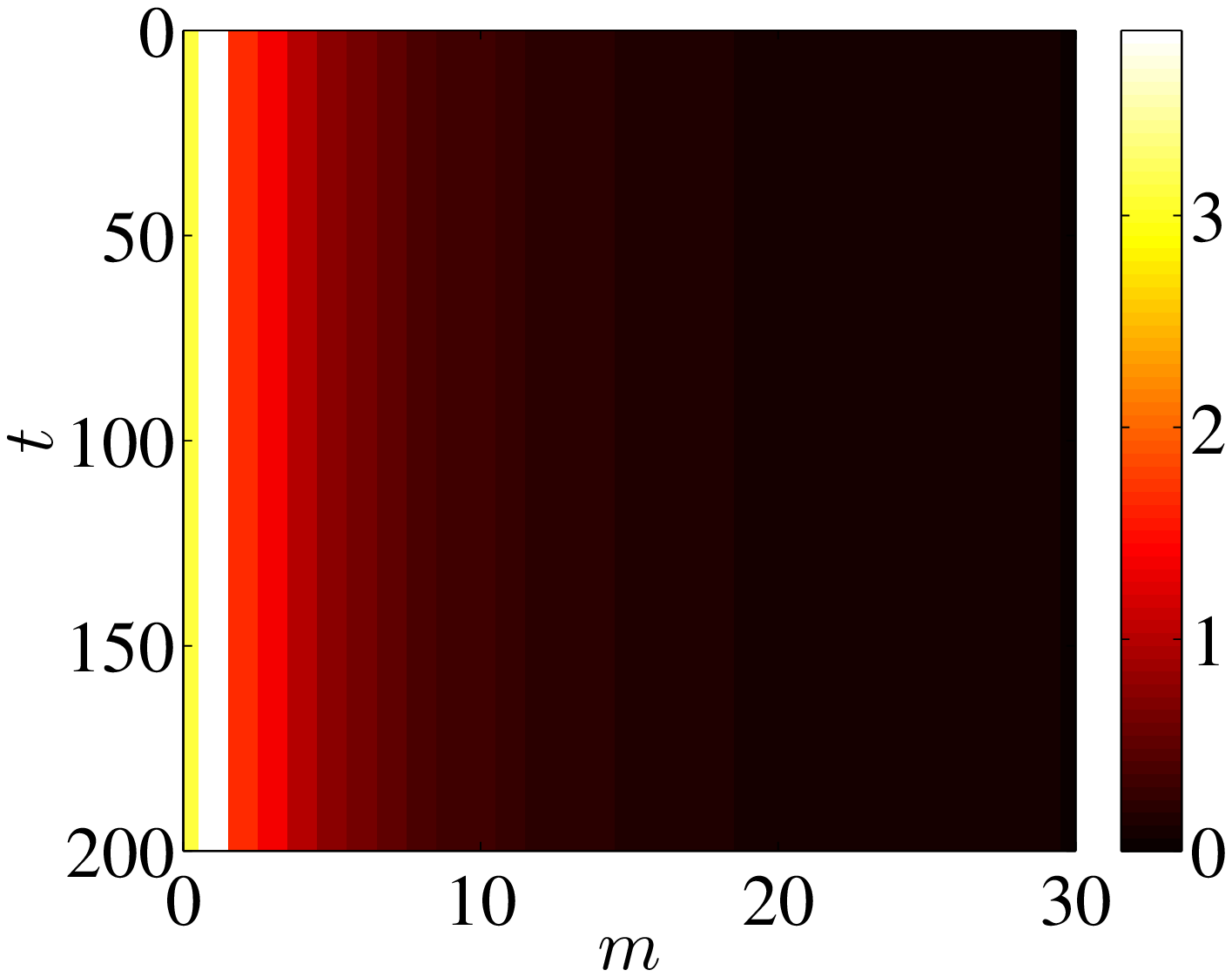}
}
\subfigure[][]{\hspace{-0.5cm}
\includegraphics[height=.18\textheight, angle =0]{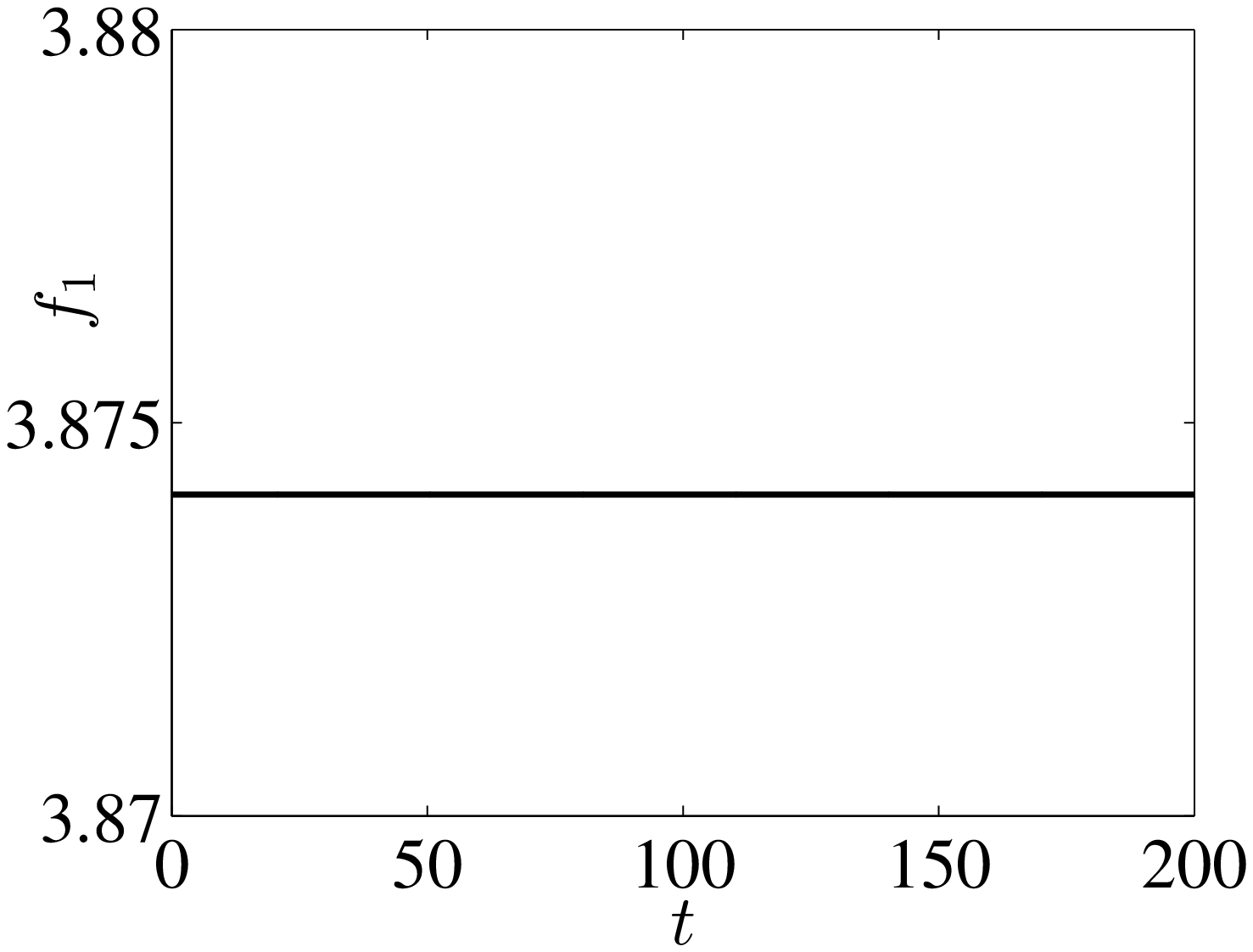}
}
\subfigure[][]{\hspace{-0.5cm}
\includegraphics[height=.18\textheight, angle =0]{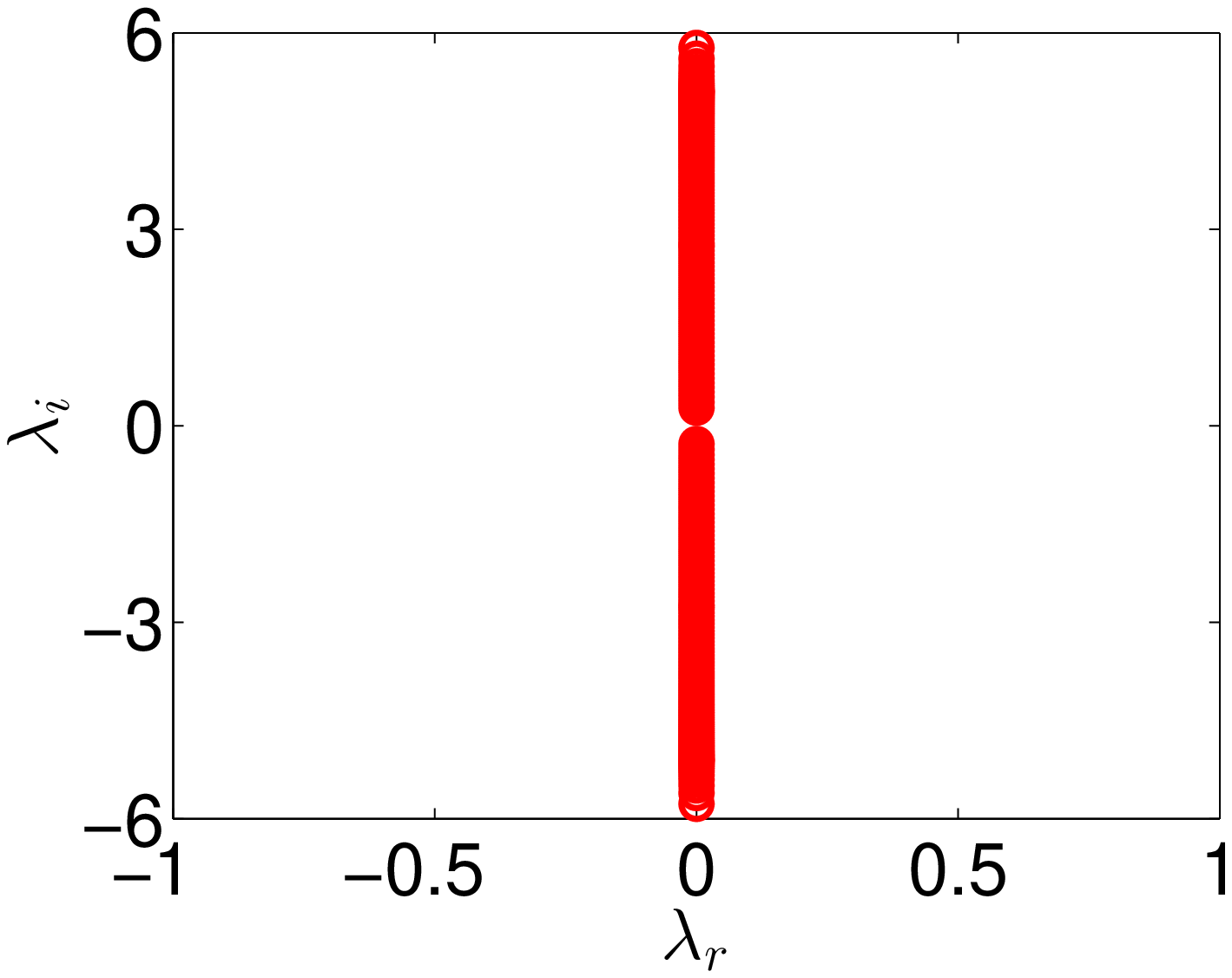}
}
}
\mbox{\hspace{0.0cm}
\subfigure[][]{\hspace{-0.5cm}
\includegraphics[height=.18\textheight, angle =0]{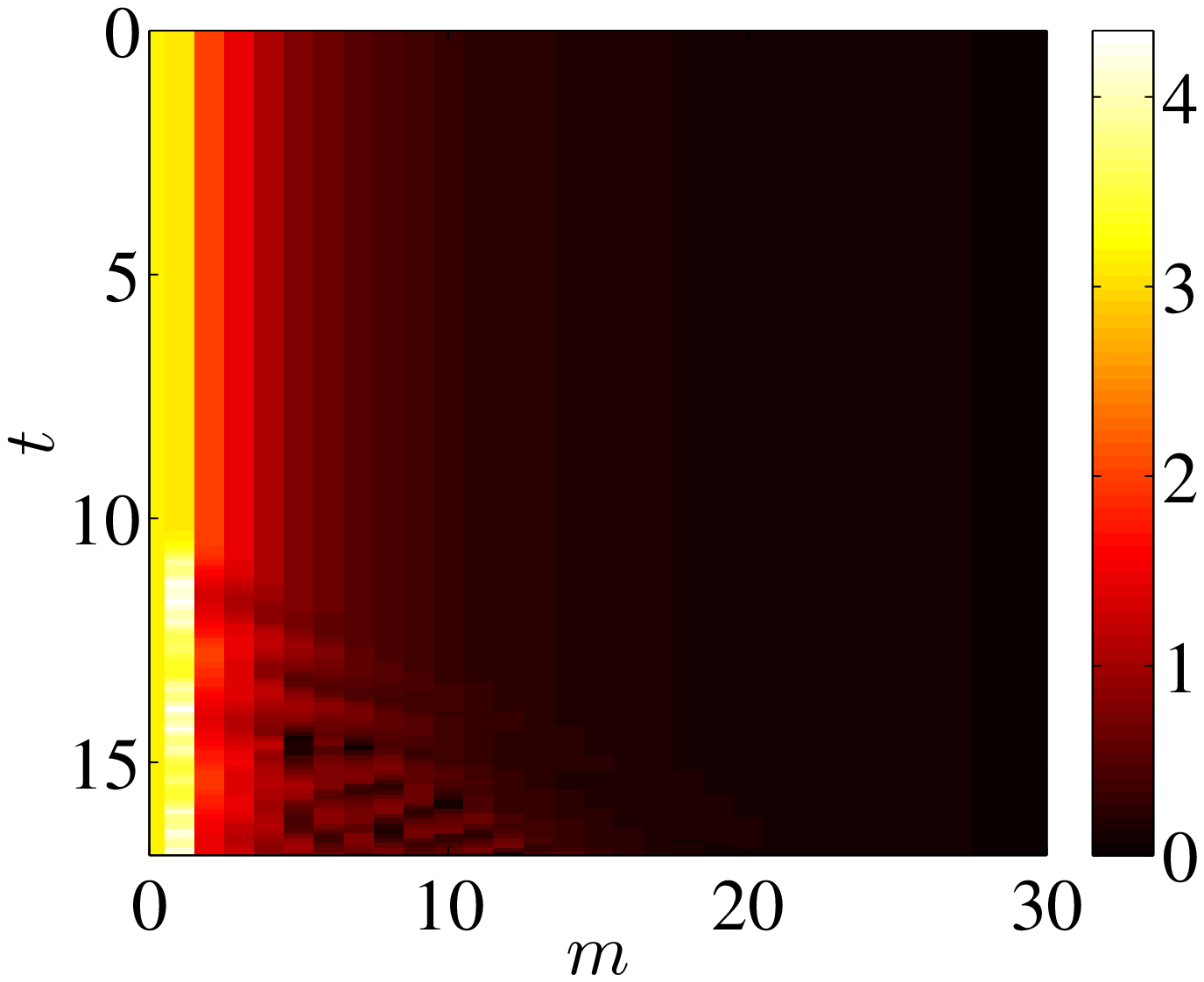}
}
\subfigure[][]{\hspace{-0.5cm}
\includegraphics[height=.18\textheight, angle =0]{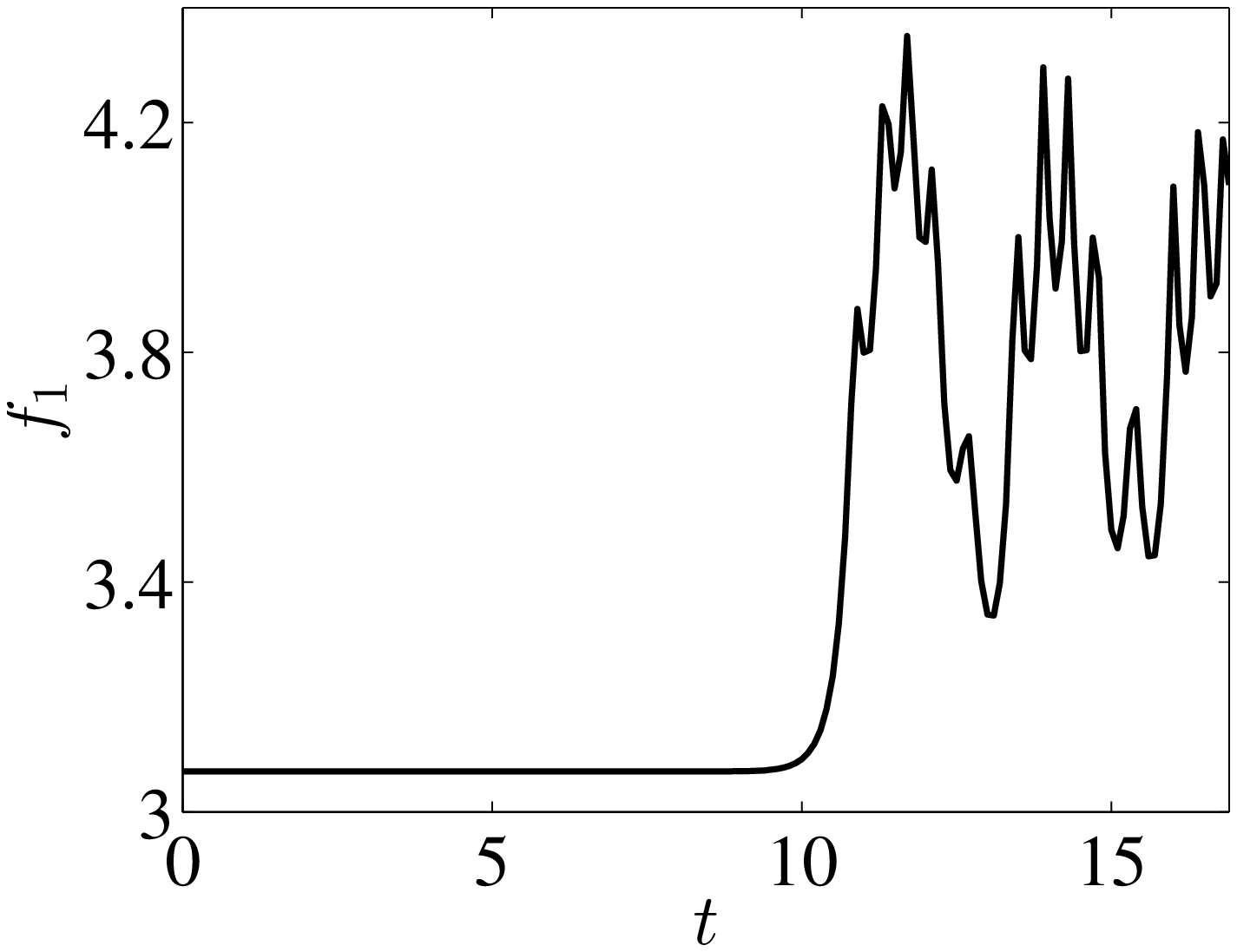}
}
\subfigure[][]{\hspace{-0.5cm}
\includegraphics[height=.18\textheight, angle =0]{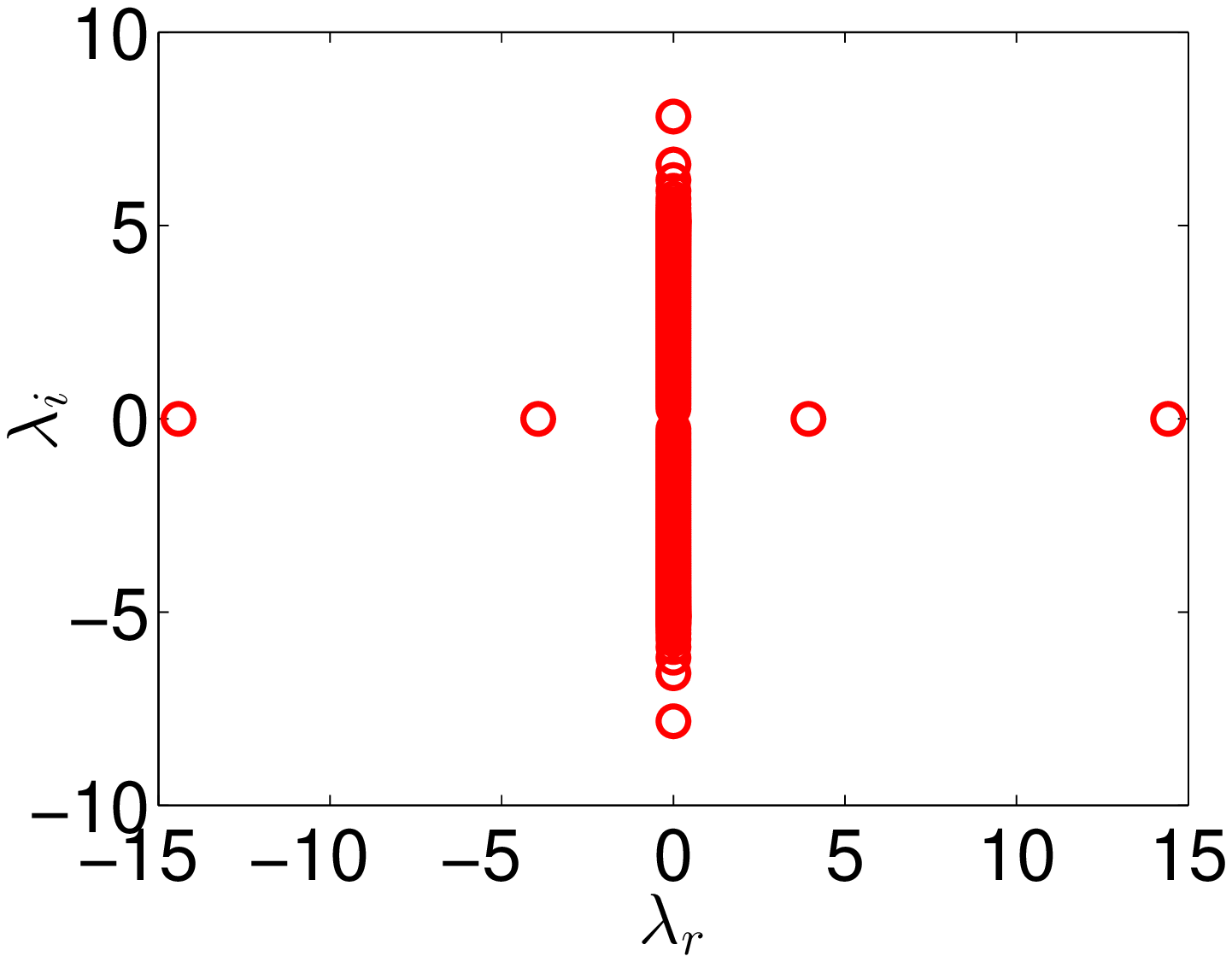}
}
}
\end{center}
\caption{
Same as Fig.~\ref{fig6} but for $h=0.4$. Spatiotemporal 
evolution of the steady-state solutions of panels (c) and (d) of Figure~\ref{fig3}
and associated spectra are shown at the top and bottom panels, respectively. 
Again, the left, middle and right panels are the same as before.
\label{fig7}
}
\end{figure}

Next, we study  the dynamics of the  BPS skyrmions for $h=0.3$ and $h=0.4$ while $\alpha=4$
and $\alpha=5$  and investigate their stability.
It is now of crucial importance to highlight that the time-dependent
ODEs
%~(\eqref{eq_m=1}) and
\eqref{eq_m>1} require careful handling from the 
numerical computations' point of view. 
Specifically, the second order in-time
system involves a denominator of the form of $\sin^{4}(\frac{f_{+}+f}{2})$ 
which for large $m$, i.e., far from the origin, becomes zero due to
the fact that the discrete profiles asymptote to zero.
{This leads to significant complications in the numerical computations,}
and a special treatment of such terms is needed, in order 
to avoid overflow in the computations. To overcome this issue, we impose an 
artificial cut-off to the discrete profiles, that is, we 
truncate the computational domain from the original into a new one such that
the value of the profile function at the next-to-last site is of the order
of $10^{-4}$. For consistency, the same homogeneous boundary condition is
employed
and the ``new'', free-from-overflow profile is again a solution to the steady
version of equations
%.~(\eqref{eq_m=1}) and~ 
\eqref{eq_m>1}.
{In this way, a stability analysis can be carried out for such 
steady-state profiles as follows. The perturbation ansatz of the 
form of}
\begin{equation}
f=f^0 + \varepsilon\,e^{\lambda t} w, \quad (\varepsilon \ll 1), 
\end{equation}
{is introduced with $f^{0}$ being a steady state, $w\doteq w(mh)$, 
and at order $\varepsilon$, Eq.~\eqref{eq_m>1} results in an eigenvalue problem with $\left(\lambda,w\right)$ 
representing the corresponding eigenvalue and eigenvector, respectively. 
If any of the eigenvalues $\lambda=\lambda_{r}+i \lambda_{i}$ has a
positive
real part, the underlying steady state is deemed to be unstable
(on the other hand, marginal stability arises only when all the eigenvalues
are found to be imaginary, i.e., correspond to small oscillations
around the equilibrium).}
Results for the dynamics of BPS skyrmions for $\alpha=4$ and 
$\alpha=5$ with $\dot{f}_{m}=0$ ($m>1$, i.e., in the absence of initial speed)
{as well as associated spectra} 
are shown in Figures \ref{fig4} and \ref{fig5}; \ref{fig6} and \ref{fig7}, 
respectively. These case examples are performed for the points
(b), (e) and (c), (d) in each branch (and for both $\alpha=4$
and $\alpha=5$) i.e., for the cases where
discreteness plays a more pronounced role in the results. 

It can be discerned from the Figures \ref{fig4}-\ref{fig7}
that the BPS skyrmions are generally deemed to be unstable
{and this is also confirmed by computing the associated
spectra of the solutions (see, the right panels therein)}. 
These instabilities are either manifested via a drastic (localized)
amplitude decay as in Figures \ref{fig4}(a)-(b) and Figures \ref{fig6}(c)-(d);
or through the emission of radiative wavepackets
as in Figures \ref{fig4}(c)-(d) and Figures \ref{fig5}(c)-(d).
They may also lead to oscillatory dynamics such as those
observed in Figures \ref{fig5}(a)-(b) and Figures \ref{fig7}(c)-(d).
However, it is important to note that in the case of $\alpha=5$ 
(Figures \ref{fig6} and \ref{fig7}) for $h=0.3$ and $h=0.4$ (depicted
in panels (b) of the respective figures), the
discrete BPS skyrmions appear to be long-lived ones over a wide
time window (see the range of the $x$-axis). 
{This is also corroborated by our linear stability analysis 
results for these particular cases since all the eigenvalues $\lambda$ 
are sitting on the imaginary axis, thus suggesting that the pertaining 
waveforms are (indeed) stable}. Hence, these solutions are promising for
a discrete realization of BPS skyrmionic structures.

\section{Conclusions and Future Challenges}

In the present work, we have revisited the topic of discrete
skyrmions in the context of the so-called BPS model.
We have provided a discretization motivated by the
enforcement of the Bogomolny bound of the model.
This results in a Hamiltonian model that is, however,
highly nonlinear in that the kinetic energy term
is multiplied by a sinusoidal function of the field.
We have discussed the numerous nontrivial complications/challenges
that such a feature presents from a numerical perspective,
{as well as a possible way to overcome them}.
On the one hand, it is not possible at the present stage
to conduct a systematic spectral stability analysis of
these solutions at least in the form involving their
vanishing tails (given the relevant small denominators
that arise in the eigenvalue computation). A similar
problem renders rather difficult the examination of
the dynamical evolution of the solutions. Nevertheless,
utilizing a suitable truncation, we have % been able to monitor
{not only been able to monitor}
the waveforms in direct numerical simulations, {but  also to
corroborate our findings in the realm of linear stability analysis
in the cases reported in this work}. %The latter
{Our results}
have revealed that the identified solutions are typically
unstable. However, for larger values of the potential
parameter $\alpha$, it is possible to find long-lived case
examples of the relevant states, which are promising candidates
for discrete BPS skyrmions.

Naturally, there are numerous open questions that emerge from this
study. One such involves the fact that the higher values of $\alpha$
considered here appear to have a better chance to lead to long-lived
states. Hence, it would be interesting to find out if for sufficiently
large $\alpha$'s the solution becomes generically (potentially) robust.
Another question is that of the spectral stability: is it possible
(perhaps via tricks like the one of domain truncation utilized here)
to obtain meaningful information about the spectral stability of
these solutions? {In the present work, we have partially
answered this question,} although it would be useful to explore that systematically
%If so, it would be useful to explore that systematically
for the branches presented herein. Finally, here we have studied the
radial problem for the BPS Skyrme model case. It would be relevant
to explore the effect of azimuthal perturbations and the nature
of their impact on the stability and dynamics of the considered
discrete skyrmion states. Such studies will be considered
in future publications. 

\section{Acknowledgments}
M.A acknowledges  support from FP7, Marie Curie Actions, People, International Research Staff Exchange Scheme (IRSES-606096).
T.I. acknowledges  support from FP7, Marie Curie Actions, People, International Research Staff Exchange Scheme (IRSES-606096); and  from The Hellenic Ministry of Education: Education and Lifelong Learning Affairs, and European Social Fund: NSRF 2007-2013, Aristeia (Excellence) II (TS-3647).

\end{document}